\documentclass{aa}
\usepackage[utf8]{inputenc}
\usepackage{graphicx}
\usepackage{comment}
\newcommand{\vect}[1]{\boldsymbol{#1}}
\renewcommand{\thesubsection}{\arabic{subsection}}
\usepackage{makecell}
\usepackage{hyperref}
\usepackage{verbatim}
\usepackage{pbox}
\usepackage{physics}
\usepackage{amsmath}
\let\oldAA\AA
\renewcommand{\AA}{\text{\normalfont\oldAA}}

\begin{document}

\title{Beyond binarity in A stars \\ II. Disentangling the four stars in the vicinity of the triple \\ HIP 87813 within the quintuple system HJ2814 \thanks{Based on observations collected at the European Southern Observatory, Chile, Program IDs 075.D-0342, 077.D-0164, 093.D-0346, 105.20RL.001}}
\subtitle{} 

\author{Idel Waisberg\inst{1}, Ygal Klein\inst{1} \and Boaz Katz\inst{1}} 
\institute{\weizmann} 

\newcommand{\weizmann}{Department of Particle Physics and Astrophysics, Weizmann Institute of Science, Rehovot 76100, Israel; 
\email{idel.waisberg@weizmann.ac.il}}

\abstract{A-stars are the progenitors of about half of the white dwarfs (WDs) that currently exist. The connection between the multiplicity of A-stars and that of WDs is not known and both multiplicities are still poorly explored. We are in the process of obtaining tight constraints on a sample of 108 southern A-type stars that are part of the nearby VAST sample \citep{DeRosa14} by conducting near-infrared interferometric follow-up observations to the (twenty) stars among them which have large $Gaia$-$Hipparcos$ accelerations. In this paper, we combine spectroscopy, adaptive optics imaging, NIR interferometry and $Gaia$-$Hipparcos$ astrometry in order to disentangle the stars in the complicated HIP 87813 = HJ2814A system. We show that (i) a previously discovered faint star that is separated by 2" from the A star is actually a background source; (ii) the $Gaia$-$Hipparcos$ acceleration is caused by a newly discovered $0.74 M_{\odot}$ star that was missed in previous AO images and we solve for its $P \approx 60 \text{ yrs}$ astrometric orbit; (iii) by combining previously obtained spectra we show that the A star has a very close $0.85 M_{\odot}$ companion on a 13.4-day period orbit. The radial velocity curve combined with NIR interferometry constrains its orbit allowing Kozai-Lidov oscillations in the hierarchical triple to be ruled out. The system HJ2814 is one of only about fifteen known 5+ systems with an A star primary, and will result in a system of between two to five bound WDs  within around a Hubble time.} 

% http://journals.aas.org/authors/keywords2013.html

\titlerunning{HIP 87813 is a hierarchical triple}
\authorrunning{Waisberg, Klein \& Katz}

\maketitle

%%%%%%%%%%%%%%%%%%%%%%%%%%%%%%%%%%%%%%%
% Introduction
%%%%%%%%%%%%%%%%%%%%%%%%%%%%%%%%%%%%%%%

\section{Introduction}

Multiple star systems are important in the study of star formation, interaction dynamics and the ultimate fate of stars. Systems containing one or more A stars are particularly interesting as these stars evolve within the age of the Universe to form white dwarfs (WDs), and are among the main progenitors of the WDs we see today. WDs in multiple systems are in turn believed to be the progenitors of type Ia Supernovae \citep[e.g.][]{Hillebrandt00,Maoz14}.
%, including the around 1\% of them that are believed to eventually explode as type Ia supernovae. 
However, the multiplicity of A stars is comparatively less studied then that of lower and more massive stars, and is not well known \citep[e.g.][]{Duchene13,Moe17}. 

\cite{DeRosa14} recently analyzed an adaptive optics (AO) survey of 363 (photometrically selected) A-stars within 75 pc (VAST survey), providing a complete census of most companions at separations $0.3''\lesssim\rho\lesssim10''$. Gaia EDR3 \citep{Gaia21,Lindegren21}, on the other hand, allows the detection of most main sequence (MS) companions at separations $\gtrsim5''$ by identifying common proper motion (CPM) pairs (note that the VAST survey includes a CPM search but is superseded by Gaia). Binaries at separations $\lesssim$ few AU are in principle accessible to photometric \citep[e.g.][]{Murphy18} or spectroscopic techniques with radial velocity measurements (limited to small separations due to the large width of the spectral lines of the rotating A stars), albeit such observations are highly demanding. An important knowledge gap remains for separations $0.01''\lesssim \rho \lesssim 0.3''$ \mbox{($1$ AU $\lesssim a \lesssim20$ AU for a distance $d=75$ pc)}.  Intermediate mass binaries in this separation range are, however, likely common \citep[e.g.][]{Klein17}. In fact, 4 of the closest 6 known WDs  ( Sirius B, Procyon B, 40 Eridani B and Stein 2051 B, all within 6 pc) have MS companions within \mbox{$10$ AU $< a <50$ AU}, and probably had progenitor systems consisting of an intermediate mass star in a MS-MS binary with $1$ AU $\lesssim a \lesssim 30$ AU separation. 

Missing stars in the separation range of $\sim 10$ AU may significantly affect our estimates of high multiplicity beyond binarity, with three or more components. High multiplicity is particularly challenging to infer and almost always requires a combination of different techniques -- including imaging, lightcurves, radial velocity curves, astrometry and interferometry -- which have different sensitivities and biases to orbital separation, inclination and companion contrast. While high (3+) multiplicity is very common among massive stars \citep[e.g.][]{Sana14}, the fraction of F and G dwarfs in high multiplicity (3+) systems appears to be much lower, on the order of $10\%$ \citep[e.g.][]{Tokovinin14}. This gap calls for a better understanding of the multiplicity of A stars as progenitors of systems containing one or more WDs. High multiplicity may play an important role in the evolution of the system before, during or after stellar evolution \citep[e.g.][]{Harrington68,Kiseleva98,Tokovinin06,Fabrycky07,Perets09,Toonen16,Hamers21, Gao22} and may affect the prospects of the system to result in a type Ia supernovae  \citep[e.g.][]{Thompson11,Katz12,Kushnir13}.

We are in the process of obtaining tight constraints on a volume limited sample of 108 nearby  ($40 \text{ pc} \lesssim d \lesssim 80 \text{ pc}$) southern A-type stars that are part of the VAST sample by conducting near-infrared interferometric followup observations to the (twenty) stars among them that show a large proper motion change $\Delta \text{PM} > 0.5 \text{ km}\text{ s}^{-1}$ between $Hipparcos$ and $Gaia$ \citep[based on][indicative of a nearby $M \gtrsim 1 M_{\odot}$ companion]{Brandt18,Brandt21}. In addition to the prospects of identifying the companion/s causing the acceleration (in the many cases in which a suitable companion was not detected in the AO images), the interferometry provides constraints on the 3d orientation of the close (e.g. spectroscopic) binaries, which is crucial to assess the dynamical state of the system. For a more detailed description of the survey, we refer to Paper I of this series. 

In this paper, we present new observations and analysis of one of these A stars, HIP 87813 = HJ 2814A which demonstrates the challenges and potential of combining astrometric, spectroscopic, AO and interferometric observations for studying multiplicity. HIP 87813 is an interesting example for which the $Gaia-Hipparocs$ acceleration is too large to be explained by the previously identified close companion at 2". By reanalysing previous AO images we find a new close companion which accounts for the $Gaia-Hipparcos$ acceleration and show that the previously identified close companion is a background star. By analysing previous spectroscopic observations we confirm the existence of an additional, previously suspected tight companion making HJ 2814A a triple system. By combining the imaging, spectroscopic, astrometric and new and archival interferometric observations we solve for the inner and outer orbits and put tight constraints on their orientation and possible dynamics. The entire HJ 2814 system, which includes a (previously known) common proper motion binary HJ 2814B is thus a quintuple system (see Figure \ref{fig:schematic}). This paper is structured as follows. Section 2 summarizes the properties of the system known prior to our analysis and the astrometric data from different surveys. Section 3 describes the archival and new observations and the data reduction. Section 4 presents the direct results, which are put together and further analyzed in Section 5 to reveal a consistent picture for HIP 87813. Finally, we discuss in Section 6 the ongoing dynamical processes in the system, as well as its possible formation scenarios and its future evolution. 

\section{HJ 2814 HIP 87813} 

\begin{figure}
\includegraphics[width=\columnwidth]{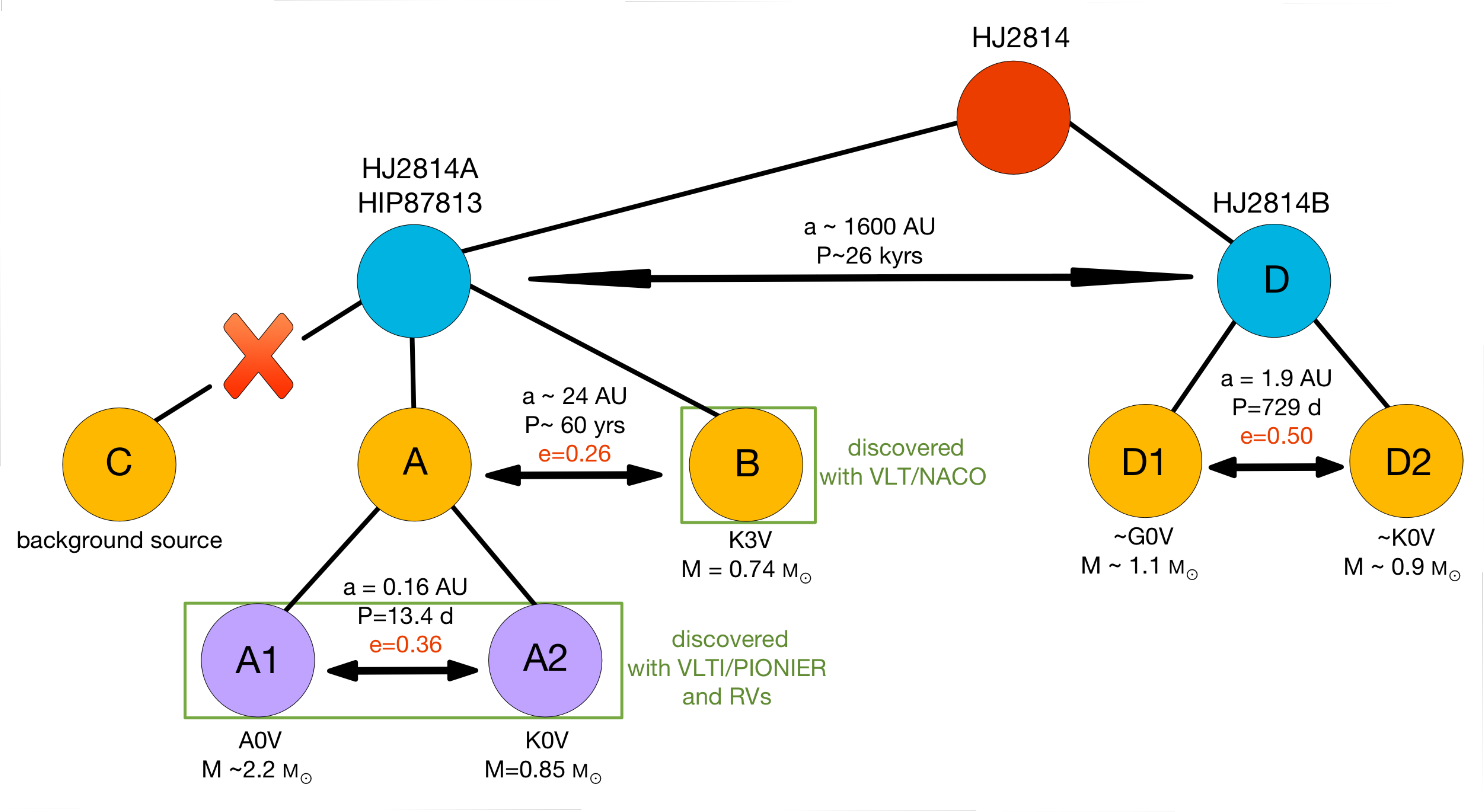}
\caption{Multiplicity schematic for HJ2814 combining the previously known information and the new discoveries in this paper.}
\label{fig:schematic}
\end{figure}

We show in Figure \ref{fig:schematic} the multiplicity structure of HJ 2014, including the previously known and the newly discovered components, which will be discussed throughout this paper. 

HIP 87813 is part of HJ 2814, which consists of two main components separated by $20.6"$, which translates to a physical projected separation of 1617 AU using the $Gaia$ eDR3 \citep{Gaia21} distance of $d=78.2 \text{ pc}$ for HIP 87813 \citep{Bailer-Jones21}. The association is extremely likely to be physical given the consistent parallaxes and close proper motions of the two objects in $Gaia$ eDR3. 

The fainter component HJ 2814B (=BD-15 4723) is a double line spectroscopic binary (D=D$_1$+D$_2$ in Figure \ref{fig:schematic}) with a period $P=729$ days and eccentricity $e=0.5$ \citep{Tokovinin19}. The color and luminosity of this component correspond to a G0V star, and the mass ratio inferred from the RV curves is $0.82$. Therefore, the system consists roughly of an early G and an early K type main sequence (MS) stars with a semi-major axis of about 1.9 AU. 

HJ 2814A, also known as HIP 87813, is the topic of this paper and is an A0-1V star. Using multi-band photometry, \cite{Zorec12} estimated its fundamental parameters: effective temperature $T_{\mathrm{eff}} = 9290 \pm 150 \text{ K}$ and luminosity $L = 23.8 \pm 1.8 L_{\odot}$ ($R = 1.9 \pm 0.1 R_{\odot}$). Using evolutionary tracks on the HR diagram they then estimated a mass $M = 2.14 \pm 0.03 M_{\odot}$ and a fractional age on the main sequence $t/t_{MS} = 0.276 \pm 0.086$ (corresponding to an age around $t \sim 400 \text{ Myrs}$). \cite{DeRosa14} also estimated a mass and age for the star by comparing the NIR K band magnitude and the $V-K$ color with theoretical isochrones, finding similar values of $M=2.11 M_{\odot}$ and $t=350 \text{ Myrs}$. 
The projected rotational velocity of the A star is reported as $v \sin i = 52 \text{ km/s}$ in \cite{Glebocki05} and $v \sin i = 60 \text{ km/s}$ in \cite{Royer07}. 

A probable binary nature for this star was noted in \cite{Tokovinin19} based on three different radial velocities reported in \cite{Nordstrom85} and on a large proper motion change between Hipparcos and $Gaia$ DR2 \citep{Gaia18}. Based on this, the system HJ 2814 is classified as a quadruple in the most recent version of the Multiple Star Catalogue \citep{Tokovinin18}. The fact that the radial velocities changed by $30 \text{ km}\text{ s}^{-1}$ over four days, however, suggests a short-period orbit that cannot explain the proper motion change, and already hint towards at least a triple star nature for HIP 87813. 

Furthermore, \cite{DeRosa14} reported the discovery of a faint companion to the A star at a projected separation $\rho=1.88"$ based on archival AO observations with VLT/NACO. From the contrast in K band magnitude $\Delta K = 6.66$ and assuming the stars are physically related, they estimated the companion mass to be $0.17 M_{\odot}$, corresponding to a mid to late M dwarf. This putative companion, however, can explain neither the radial velocity nor the proper motion changes of the A star.

\subsection{Astrometry of HIP 87813}

In table \ref{table:PM} we collect the proper motions of HIP 87813 as given by the FOCAT-S catalogue \citep{Bystrov94}, $Hipparcos$ \citep{vanLeeuwen07}, $Gaia$ DR2 \citep{Gaia18} and $Gaia$ eDR3 \citep{Gaia21}, with mean epochs J1982.492, J1991.25, J2015.5 and J2016.0 respectively. The FOCAT-S proper motion is based on two position measurements but the individual epochs are not reported. The $Hipparcos$, $Gaia$ DR2 and $Gaia$ eDR3 positions and proper motions are based on 82, 147 and 167 measurements made over a few years. We also show the ``effective'' proper motion based on the positional difference between $Gaia$ (DR2) and $Hipparcos$ \citep[e.g.][]{Brandt18,Brandt21}, which is much more precise than the individual proper motions. These values were translated to $\text{km} \text{ s}^{-1}$ using a geometric distance $d = 78.2 \text{ pc}$ based on the $Gaia$ eDR3 parallax \citep{Bailer-Jones21}. 

\begin{table*}[t]
\centering
\caption{\label{table:PM} Proper motion measuremens of HIP 87813.}
\begin{tabular}{cccc}
\hline \hline
source & epoch & \shortstack{pmra\\($\text{mas}\text{ yr}^{-1}$)\\($\text{km}\text{ s}^{-1}$)} & \shortstack{pmdec\\($\text{mas}\text{ yr}^{-1}$)\\($\text{km}\text{ s}^{-1}$)} \\[0.3cm]
FOCAT-S & J1982.492 & \shortstack{$-5\pm1$\\$-1.9\pm0.4$} & \shortstack{$-63\pm5$\\$-23.4\pm1.9$} \\[0.3cm]
$Hipparcos$ & J1991.25 & \shortstack{$-9.77\pm0.50$\\$-3.62\pm0.19$} & \shortstack{$-65.32\pm0.30$\\$-24.23\pm0.11$} \\[0.3cm]
$Gaia$ DR2 & J2015.5 & \shortstack{$-1.44\pm0.20$\\$-0.53\pm0.07$} & \shortstack{$-72.79\pm0.16$\\$-27.00\pm0.06$} \\[0.3cm]
$Gaia$ eDR3 & J2016.0 & \shortstack{$-1.60\pm0.11$\\$-0.59 \pm 0.04$} & \shortstack{$-72.68\pm0.07$\\$-26.96 \pm 0.03$} \\[0.3cm]
$Gaia$ DR2 -$Hipparcos$ & - & \shortstack{$-4.44\pm0.02$\\$-1.647\pm0.009$} & \shortstack{$-71.36\pm0.02$\\$-26.469\pm0.007$} \\[0.3cm]
\hline
\end{tabular}
\end{table*}

There is a significant change in both the short-term proper motions measured by $Hipparcos$ and $Gaia$ DR2 (|$\Delta \mathrm{PM}| = \sqrt{(\mathrm{PM}_{\mathrm{RA}})^2 + (\mathrm{PM}_{\mathrm{DEC}})^2 } = 3.90 \pm 0.33 \text{ km}\text{ s}^{-1}$) as well as between $Gaia$ DR2 and the effective proper motion based on the positional difference between $Gaia$ DR2 and $Hipparcos$ (|$\Delta \mathrm{PM}| = 1.24 \pm 0.07 \text{ km}\text{ s}^{-1}$). On the other hand, there is no significant proper motion change between $Gaia$ DR2 and eDR3 (|$\Delta \mathrm{PM}| = 0.07 \pm 0.08 \text{ km}\text{ s}^{-1}$). This suggests that the acceleration is caused by an orbit with a period $P \gg 1 \text{ yr}$. 

On the other hand, $Gaia$ eDR3 reports both a high astrometric noise of $0.62 \text{ mas}$ (with a significance of 277) and a high RUWE (Renormalized Unit Weight Error) parameter \citep{Lindegren21} of 2.84 for HIP87813. These are both indications that the single-star solution is very poor and that the star is possibly a multiple \citep[e.g.][]{Belokurov20}. Since there is no change in proper motion between DR2 and eDR3, the astrometric noise is likely to be caused by a very short period ($P \ll 1 \text{ yr}$) binary whose orbit averages out in the measurement of the $Gaia$ proper motions, rather than by a residual from the acceleration caused by the long period binary.  

The astrometric data therefore provides further evidence that the A star in HIP 87813 is actually at least a triple system, 
the nature of which can be elucidated with the radial velocity, imaging and interferometric observations that will be presented below.

\section{Observations and data reduction}

\subsection{Spectra} 

We collected FEROS spectra of HIP87813 from the ESO Phase 3 archive\footnote{\url{http://archive.eso.org/wdb/wdb/adp/phase3_main/form?collection_name=FEROS}}. FEROS \citep{Kaufer99} is an Echelle spectrograph mounted on the MPG/ESO 2.2-m telescope in ESO La Silla observatory, Chile. The instrument covers the wavelength region 3500-9200$\AA$ at high spectral resolution ($R\approx48000$). The observations of HIP87813 were carried out on 2005-05-26 (8 integrations throughout 9 hours), 2005-05-27 (7 integrations throughout 7 hours), 2005-05-28 (two integrations separated by 2 hours) and 2006-06-29 (4 observations throughout 3 hours). The integration times varied between 240 and 600 seconds. We downloaded the spectra fully reduced and calibrated. We note that in the phase 3 spectra the wavelengths are in air and are already corrected to the heliocentric frame.  

We also collected spectra of HIP87813 from the NASA's Infrared Telescope Facility (IRTF) archive\footnote{\url{https://irsa.ipac.caltech.edu/applications/irtf/}}, which collects and organizes observations from 2016. IRTF harbors a 3-m telescope optimized for infrared astronomy and located in Mauna Kea observatory in Hawaii. Being an A0V star, HIP 87813 was often observed as a telluric calibrator for several different programs. We downloaded the raw data and calibration files for observations with the medium resolution spectrograph SpeX \citep{Rayner03}. We retrieved the observations with the \texttt{SXD} grating (covering $0.7-2.5 \mu \text{m}$ with resolution $R\sim2000$) or, if it did not exist, with the \texttt{LXD long} grating (covering $2-5.3 \mu\text{m}$ with resolution $R\sim2500$). The total exposure times per observation are in the range of $1-10$ seconds. We reduced the data using the publicly available pipeline \texttt{Spextool} \citep{Cushing04}, which includes flat fielding, wavelength calibration, order extraction and merging. We note that the resulting IRTF spectra are in vacuum wavelengths. 

\subsection{AO imaging} 

From the ESO archive, we collected three epochs of NIR AO observations of HIP87813 taken with NAOS-CONICA \citep[NACO; ][]{Lenzen03,Rousset03} when it was mounted on the 8.2-m telescope UT4 of the Very Large Telescope (VLT) at ESO Paranal, Chile. The details of the observations are listed in Table \ref{table:NACO}. All images were obtained in the $K_s$ band with the S27 camera, which provides a 28"x28" field of view with a 27 mas $\text{pixel}^{-1}$ scale. In addition, a neutral density filter ("ND short") which reduces the flux by a factor of 80 was used in order to avoid saturation of the detector. The nearby star 2MASS 17564161-1551076 was used as a natural guide star for the AO corrections. 

\begin{table*}[t]
\centering
\caption{\label{table:NACO} Summary of archival NACO observations.}
\begin{tabular}{cccc}
\hline \hline
Date & \shortstack{DIT\\(sec)} & \shortstack{integration time\\(sec)} & \shortstack{seeing\\@ 500 nm (")} \\ [0.3cm]
2004-06-27 & 2.5 & 450 & 1.3-1.5 \\ [0.3cm]
2006-04-28 & 1 & 336 & 1.6-2.4 \\ [0.3cm]
2006-05-19 & 4 & 588 & 1.4 \\ [0.3cm]
\hline
\end{tabular}
\end{table*}

We reduced the data with the standard ESO NACO pipeline version 4.4.9 following the standard steps of dark subtraction, flat fielding using twilight sky frames and co-adding all the images. The FWHM of the resulting Point Spread Function (PSF), measured by fitting a two-dimensional Gaussian to the star, is about 110 mas in all three images.

\subsection{NIR interferometry} 

\subsubsection{VLTI/PIONIER}

\begin{figure}
 \includegraphics[width=\columnwidth]{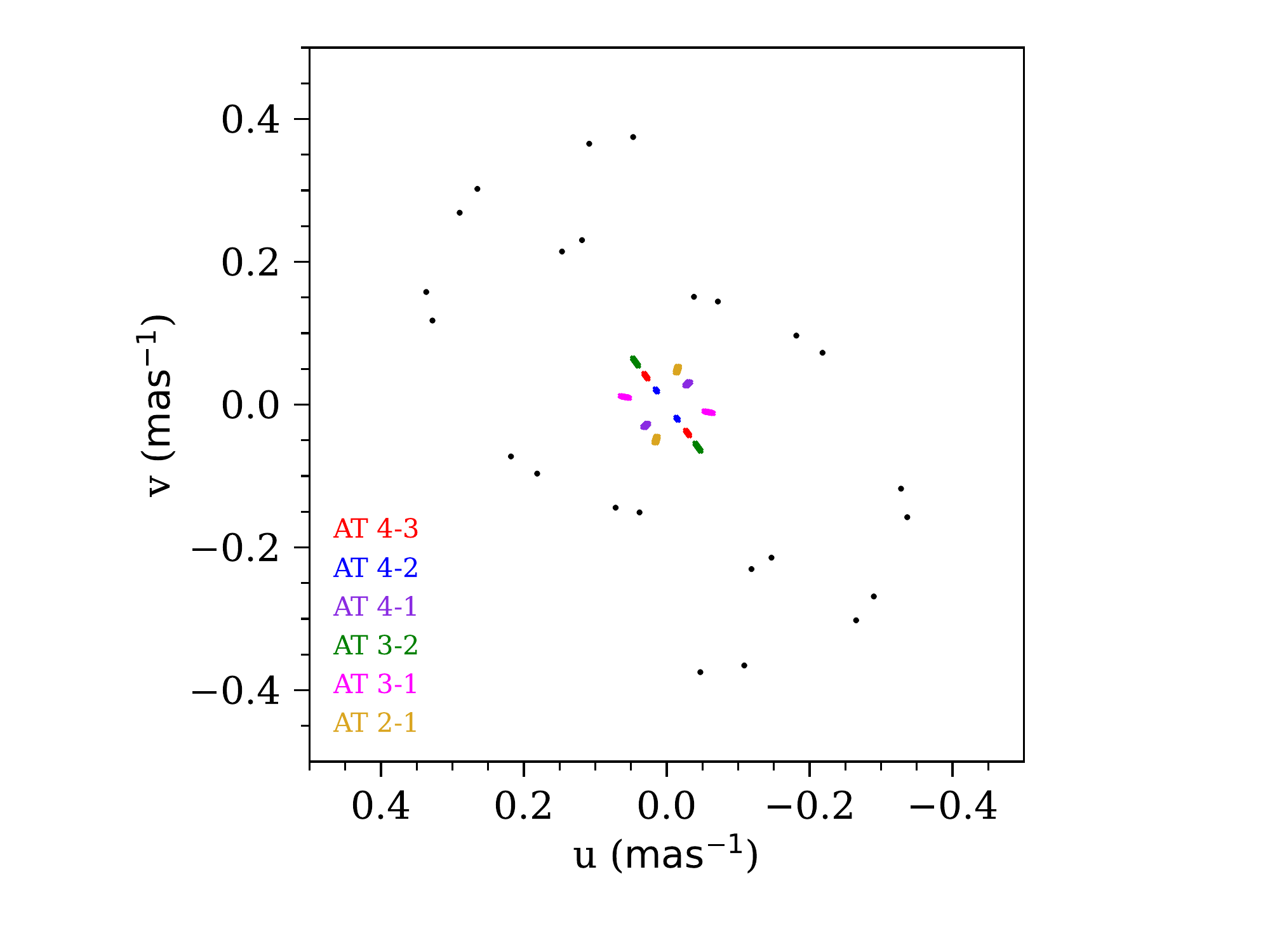}
 \caption{$uv$ coverage of VLTI observations of HIP 87813. The black points correspond to the PIONIER observations and the colorful points to the GRAVITY observations.}
 \label{fig:uv}
\end{figure}

HIP 87813 was observed with the beam combiner instrument PIONIER \citep{LeBouquin11} at the Very Large Telescope Interferometer (VLTI) in ESO Paranal on August 26th, 2014, as a calibrator star for the program 093.D-0346 (PI: Schoeller). It was observed in the one spectral channel mode ("H FREE"), with central wavelength $1.681 \text{ $\mu$m}$ and bandwidth $0.245 \text{ $\mu$m}$, in two different sets of 5 recordings separated by 1h30min within the night with average seeing conditions. Each recording is a 30s exposure containing 100 inteferograms. Each set provides six squared visibilities ($\mathrm{V}^2$) for each of the six baselines and four closure phases ($\phi_3$) for each of the four telescope triangles. Therefore, the entire observations comprise twelve squared visibilities and eight closure phases. The observations were made using the four 1.8-m Auxiliary Telescopes (ATs) in the station configuration A1-G1-J3-K0, which provided a maximum projected baseline length of 140 meters, corresponding to an angular resolution of $2.5 \text{ mas}$ (we note that structures up to a few times smaller than the maximum resolution can be partially resolved). The resulting $uv$-coverage is shown in Figure \ref{fig:uv}. A summary of the observations is reported in Table \ref{table:obs}. 

\begin{table*}[t]
\centering
\caption{\label{table:obs} Summary of archival PIONIER and obtained GRAVITY interferometric observations.}
\begin{tabular}{cccccc}
\hline \hline
instrument & \shortstack{date\\time(UTC)} & \shortstack{avg.seeing\\@ 500 nm (")} & \shortstack{avg.coherence time\\@ 500 nm (ms)} & \shortstack{AT configuration} & calibrator \\ [0.3cm]
PIONIER & \shortstack{2014-08-26\\00:30-00:32} & 0.75 & 2 & A1-G1-J3-K0 & HD 158352 \\ [0.3cm]
PIONIER & \shortstack{2014-08-26\\02:09-02:12} & 1.00 & 1.5 & A1-G1-J3-K0 & HD 158352 \\ [0.3cm]
GRAVITY & \shortstack{2021-09-02\\02:17-02:36} & 0.74 & 3.6 & A0-B2-D0-C1 & HD 163449 \\ [0.3cm]
\hline
\end{tabular}
\end{table*}

We note that HIP 87813 was observed as an interferometric calibrator which, in principle, would make it impossible to extract scientific information from its observations if they were isolated (we note that the fact, as will be shown below, that HIP 87813 is not a single star would make it innappropriate as an interferometric calibrator, a misfortune that sometimes happens in optical/NIR interferometry). Luckily, however, the observations were part of a very long sequence of homogeneous PIONIER observations including several targets throughout the night; therefore, it was possible to use another (true) calibrator star and treat HIP 87813 as a scientific target. In this case, we chose the calibrator HD 158352, which was observed within 15 min of HIP 87813 and with the same instrumental setup, and whose details are shown in Table \ref{table:calibrators}. Altough the on-sky distance between the two stars is larger ($\approx 17.5^{\degr}$) than would typically be selected for an interferometric calibrator ($\lesssim 3^{\degr}$), we found that the systematic calibration errors -- which for canonical observations with PIONIER are around $3\%$ for $\mathrm{V}^2$ and $2^{\degr}$ for $\phi_3$ \citep{LeBouquin11} -- are still small enough to allow the data to be useful, albeit with a larger error in the calibrated squared visibilities.

\begin{table}[t]
\centering
\caption{\label{table:calibrators} Interferometric calibrators.}
\begin{tabular}{cccccc}
\hline \hline
calibrator & \shortstack{spectral\\type} & V & H & K & $\theta_{H/K}$ (mas) \\[0.3cm]
HD 158352 & A8V & 5.4 & 4.9 & 4.8 & $0.41 \pm 0.01$\tablefootmark{a} \\[0.3cm]
HD 163449 & K0III & 7.6 & 4.9 & 4.7 & $0.56 \pm 0.01$\tablefootmark{a} \\[0.3cm]
\hline
\end{tabular}
\tablefoot{\newline
\tablefoottext{a}{From the JMMC Stellar Diameter Catalog (JSDC) v2 \citep{Bourges17}.}
}
\end{table}

We downloaded the data from the ESO archive, including the calibration files (dark and kappa matrix frames), and reduced the data using the default settings in the PIONIER data reduction software \texttt{pndrs} v3.94 \citep{LeBouquin11} downloaded from the Jean-Marie Mariotti
Center (JMMC) website. The only additional step needed was to change the label of the observations of HIP 87813 from calibrator to science for the final calibration step. 

\subsubsection{VLTI/GRAVITY}

HIP 87813 was observed with the beam combiner instrument GRAVITY \citep{GRAVITY17} on the VLTI on September 2nd, 2021, as one of the targets in our program to investigate the multiplicity of A stars that show large proper motion changes between $Hipparcos$ and $Gaia$. GRAVITY operates in the NIR K band and the observations were made in single field mode: half of the light from the object was directed to the fringe tracker (FT), which operates in low spectral resolution mode ($R\approx22$) in order to track and stabilize the fringes, while the other half was directed to the science (SC) channel, where coherent integrations of 30s were made in the high spectral resolution ($R\approx4000$) mode. The observations were done in combined polarization mode and made use of the adaptive optics (AO) system NAOMI \citep{Woillez19} using the optical light from the object itself. The total integration time on source was 16 min. The observations provided spectrally resolved squared visibilities and closure phases across the K band (1.97-2.40$\mu$m). The observations used the four ATs in the (``small'') configuration A0-B2-D0-C1, which provided a maximum projected baseline of 33 m, corresponding to an angular resolution of 12 mas. The resulting $uv$-coverage is shown in Figure \ref{fig:uv}. A summary of the observations is shown in Table \ref{table:obs}. GRAVITY also records an $80 \text{ mas} \text{ pixel}^{-1}$ H band image of 4" around the target every second for each telescope with its acquisition camera. 

The data were reduced with the GRAVITY instrument pipeline v1.4.0 \citep{Lapeyrere14} downloaded from the ESO website. The star HD 163449 was used as an interferometric calibrator, and its properties are listed in Table \ref{table:calibrators}. 

\section{Data analysis and direct results}

\subsection{Radial velocity measurements}

We searched the Vizier online data catalogues and found four radial velocity measurements of HIP 87813: three in \cite{Nordstrom85} (one from 1975 and two from 1977 separated by 4 days) and one in \cite{Grenier99} (from 1994). These values already clearly show that it is an RV variable and are listed in Table \ref{table:RV_hist}. 

The FEROS spectra cover the blue optical region where strong and narrow metallic lines are particularly suitable for radial velocity measurements of A stars. We measure a radial velocity in each individual spectra by fitting a Gaussian profile to the strong and isolated CaII K line (rest wavelength $3933.663 \AA$ in air) using the \texttt{splot} routine in IRAF. The errors are calculated through a bootstrapping routine, with the flux errors estimated from the RMS in the adjacent continuum region. The resulting statistical RV errors vary from $20-260 \text{ m} \text{ s}^{-1}$, and a clear change in RV within a single night can be seen. By finding the RMS dispersion between the eight (seven) observations in the first (second) nights after subtracting a linear fit, we find an RMS of $ 100 \text{ m} \text{ s}^{-1}$, which is consistent with the estimated statistical errors. The individual RV measurements are listed in Table \ref{table:RV_feros}.

We measure the radial velocity from the IRTF spectra by fitting a Lorentzian profile to the Pa$\beta$ line (rest wavelength $12821.39 \AA$ in vacuum) or to the Br$\gamma$ line (rest wavelength $21660.87 \AA$) for the \texttt{SXD} and \texttt{LXD long} gratings, respectively. The statistical errors in the radial velocity measurements are on the order of $5 \text{ km} \text{ s}^{-1}$; however, one might expect the errors to be systematic-dominated in this case because of the relatively low spectral resolution and the broad hydrogen lines. We use narrow telluric lines near the hydrogen lines in order to refine the wavelength calibration. Specifically, we use the telluric lines at $12688.50 \AA$ and $22001.59 \AA$ for the Pa$\beta$ and Br$\gamma$ lines, respectively, with their rest wavelength measured from the atmospheric transmission profile at spectral resolution $R=2000$ provided in \texttt{Spextool}. We find corrections that vary between $-50$ and $+68 \text{ km} \text{ s}^{-1}$, and which are therefore crucial to be taken into account. Finally, we correct the velocities to the heliocentric frame using the function \texttt{rvcorrect} in IRAF. The individual RV measurements are listed in Table \ref{table:RV_spex}.

The GRAVITY data taken at $R=4000$ provides a K band spectrum from which a radial velocity can also be extracted. We averaged the spectrum from the four telescopes and for the four files, and found a velocity $v_z = -57 \pm 10 \text{ km}\text{ s}^{-1}$ by fitting a Lorenztian profile to the Br$\gamma$ line, and refining the wavelength calibration using the nearby telluric line as above. For reference, we note that the GRAVITY spectra returned by the pipeline are in vacuum wavelengths and not corrected to the heliocentric frame. 

\subsection{VLT/NACO images} 

\begin{figure*}
 \includegraphics[width=2\columnwidth]{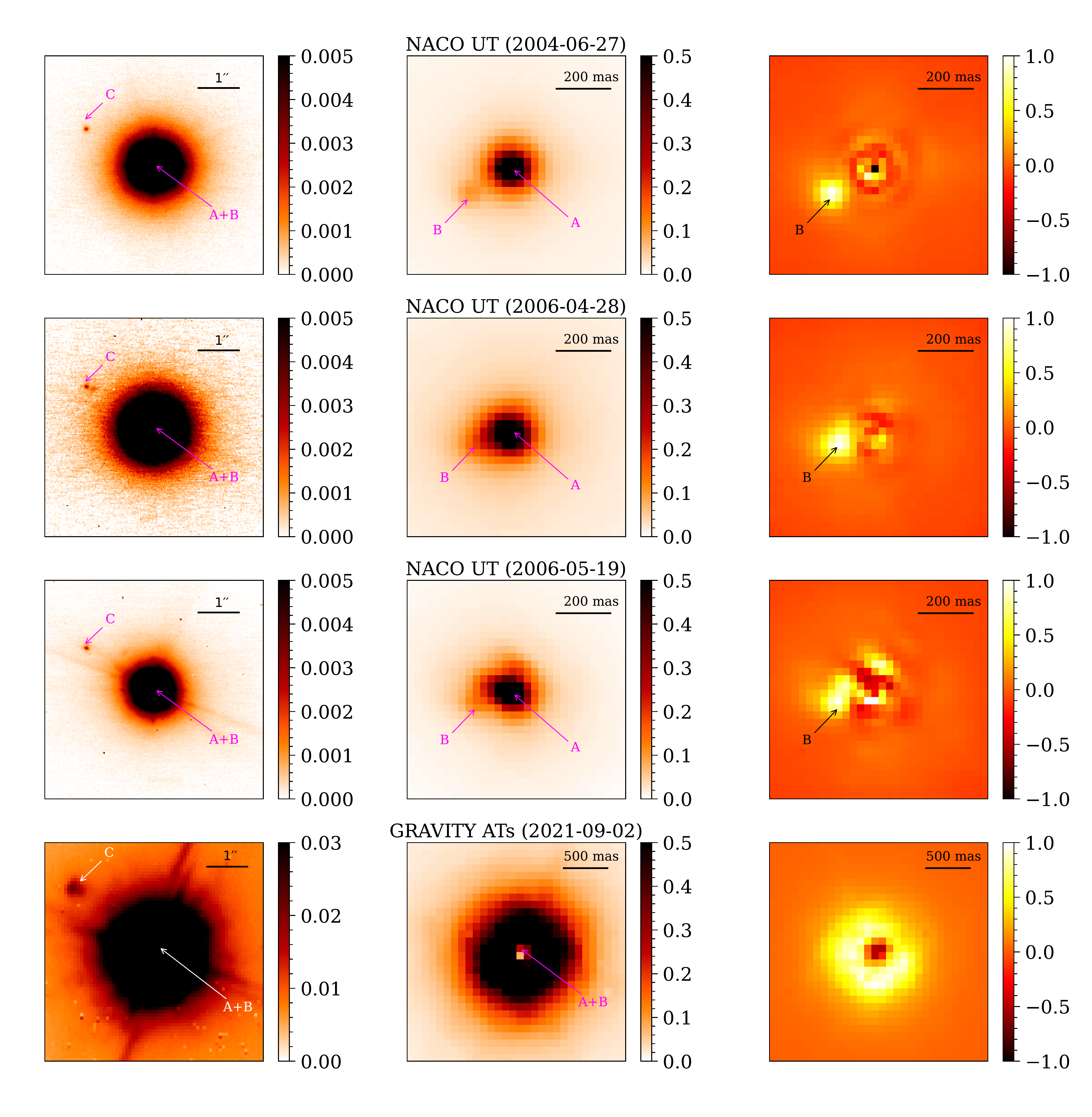}
 \caption{NIR images of HIP 87813. The first three rows show the NACO observations in K band and the last row shows the image of the acquisition camera of GRAVITY in H band. The left column highlights the faint wide companion (C) reported in \cite{DeRosa14}. The second column highlights the closer companion (B) which we have discovered, while the third column shows the residual image after subtracting the best fit component for the brighter source (A). Note that component B could not be resolved in the GRAVITY image, in which the A component has an effective PSF of FWHM 1" due to saturation.}
 \label{fig:NACO}
\end{figure*}

Direct inspection of the AO NACO images showed not only the very faint star at 2" separation detected by \cite{DeRosa14}, but also another brighter companion in the halo of the A star. We speculate that the latter was missed by \cite{DeRosa14} simply because the right contrast is needed in the image in order to make it clearly visible. From now on, the we will refer to the A star, the close companion and the wider companion as A, B and C, respectively (see Figure \ref{fig:schematic}). We plot the NACO images in Figure \ref{fig:NACO}, with the left column highlighting the C component to the northeast and the middle column the B component to the (south)west. The latter has clearly moved northwards in around two years. 

From the images we can measure the flux and position of B and C relative to A. While this is straightforward for component C, for component B PSF fitting is necessary because it is blended or sits in the halo of the much brighter A component. We experimented with a few different analytical models for the PSF, and found the best results with an elliptical Moffat profile \citep{Moffat69}. In this model the intensity of the source is

\begin{align}
I (x,y) = (1 + A (x-x_0)^2 + B (y-y_0)^2 \\ + C (x-x_0) (y-y_0))^{-\beta}
\end{align}

\noindent where $(x_0,y_0)$ is the position of the source and 

\begin{align}
A = \frac{\cos(\theta)^2}{\alpha_x^2} + \frac{\sin(\theta)^2}{\alpha_y^2} \\
B = \frac{\sin(\theta)^2}{\alpha_x^2} + \frac{\cos(\theta)^2}{\alpha_y^2} \\
C = \sin(2\theta) \left ( \frac{1}{\alpha_x^2}-\frac{1}{\alpha_y^2} \right )
\end{align}

\noindent where $\theta$ is the position angle of the ellipse and 

\begin{align}
\alpha_{x,y} = \frac{\text{FWHM}_{x,y}}{2\sqrt{2^{1/\beta}-1}}
\end{align}

\noindent Together, $\alpha_{x,y}$ and $\beta$ control the shape of the core and the wing of the PSF. In Table \ref{table:NACO_results} we report the results of the PSF deblending for each of the three NACO epochs. In the right column of Figure \ref{fig:NACO}, we plot the image with the best fit A component subtracted in order to highlight the B component. The residuals, which are particularly strong on the third image, are due to residual structure in the PSF not captured by the model. For the computation of the $\chi^2_{\mathrm{red}}$, the errors were calculated by adding the estimated background and Poisson noise in quadrature. Component B is found to have a flux $6.2 \pm 0.2 \%$ of component A (the first NACO epoch is the most reliable for the flux measurement since the blending is less severe). 

\begin{table*}[t]
\centering
\caption{\label{table:NACO_results} NACO PSF fitting results for deblending of components A and B.}
\begin{tabular}{ccccccccc}
\hline \hline
epoch & \shortstack{$\Delta$RA\\(mas)} & \shortstack{$\Delta$DEC\\(mas)} & \shortstack{$\frac{f_B}{f_A}$\\ \%} & \shortstack{$\text{FWHM}_x$\\(mas)} & \shortstack{$\text{FWHM}_y$\\(mas)} & \shortstack{$\theta$\\(deg)} & $\beta$ & $\chi^2_{\mathrm{red}}$ \\[0.3cm]
2004-06-27 & $158.5 \pm 1.6$ & $-90.7 \pm 1.4$ & $6.2 \pm 0.2$ & $3.12 \pm 0.03$ & $2.94 \pm 0.02$ & $0.5 \pm 2.4$ & $1.44 \pm 0.02$ & $0.82$ \\[0.3cm]
2006-04-28 & $124.7 \pm 1.6$ & $-35.7 \pm 1.4$ & $9.5 \pm 0.3$ & $3.46 \pm 0.03$ & $3.18 \pm 0.03$ & $-4.5 \pm 1.9$ & $1.49 \pm 0.03$ & $0.18$ \\[0.3cm]
2006-05-19 & $133.6 \pm 4.0$ & $-23.2 \pm 3.2$ & $6.6 \pm 0.5$ & $2.77 \pm 0.05$ & $3.31 \pm 0.06$ & $68.5 \pm 1.7$ & $1.33 \pm 0.05$ & $5.64$ \\[0.3cm]
\hline
\end{tabular}
\end{table*}

The separation between the C and A components was measured by finding the centroid of the C component in the residual image, and are listed in Table \ref{table:C_separation}. The flux ratio is consistent with $\Delta K = 6.7$ reported in \cite{DeRosa14}. 

\begin{table}[t]
\centering
\caption{\label{table:C_separation} Separation between components C and A for the three NACO epochs and C and A+B in the GRAVITY epoch.}
\begin{tabular}{ccc}
\hline \hline
epoch & \shortstack{$\Delta$RA\\(mas)} & \shortstack{$\Delta$DEC\\(mas)} \\ [0.3cm]
2004-06-27 & $1656.2 \pm 1.1$ & $904.5 \pm 1.1$ \\ 
2006-04-28 & $1638.4 \pm 4.6$ & $1026.0 \pm 3.3$ \\
2006-05-19 & $1644.3 \pm 1.2$ & $1033.0 \pm 0.9$ \\
2021-09-07\tablefootmark{a} & $1628 \pm 14$ & $2033 \pm 25$ \\ 
\hline
\end{tabular}
\tablefoot{\newline
\tablefoottext{a}{Separation between and C and A+B because the latter cannot be resolved in the GRAVITY image.}
}
\end{table}

\subsection{NIR interferometry} 

\subsubsection{Binary model}

For both PIONIER and GRAVITY, our basic model is that of a binary system with complex visibility \citep[e.g.][]{Waisberg19}: 

\begin{equation}
\label{eqn:binary}
V = \frac{1 + f_c f e^{-2 \pi j \vect{u}\cdot \sigma}}{1 + f_c f}
\end{equation}

\noindent where $0<f<1$ is the intrinsic binary flux ratio between the secondary and the primary in the observed band, $\vect{u}$ is the $uv$ coordinate and $\sigma$ is the binary separation vector (which we parameterize as the separation $\rho$ and the position angle PA (east of north) of the secondary relative to the primary). $f_c$ is a fiber coupling constant that is important for very wide binaries and takes into account the attenuation of sources that are far away from the center of the fiber. We model it as

\begin{equation}
f_c = e^{-|\sigma|^2/(2 \sigma_{\mathrm{fiber}}^2)}
\end{equation}

\noindent where $\sigma_{\mathrm{fiber}} = \mathrm{FWHM}_{\mathrm{fiber}}/2.355$ is the standard deviation of the fiber mode, which in the case of GRAVITY is matched to the Point Spread Function (PSF) of the telescope ($\mathrm{FWHM}_{\mathrm{fiber}} = 250 \text{ mas}$ in the case of the 1.8m ATs). Note that in Eq. \ref{eqn:binary} we assume that the fiber is centered on the primary star (which is an excellent approximation when $f \ll 1$). We also note that in Eq. \ref{eqn:binary}:

\begin{enumerate}
\item We treat the stars as point sources because their size is much smaller than the interferometer resolution. For example, the A0V primary with a radius $R\approx2R_{\odot}$ has an angular diameter $\theta \approx 0.2 \text{ mas}$ at its distance of $d=78.2 \text{ pc}$.

\item We neglect the effect of bandwidth smearing, which reduces the visibility for sources far from the center of the interferometric field of view $\mathrm{FOV}_{\mathrm{interf}} \sim 0.6 \frac{\lambda R}{B}$, where $R$ is the spectral resolution and $B$ the projected baseline. For our GRAVITY data taken at $R\approx4000$, $\mathrm{FOV}_{\mathrm{interf}} \sim 30"$ for $\lambda=2\text{ $\mu$m}$ and $B=30\text{ m}$ and this is a completely negligible effect. For the broadband PIONIER data with $R\approx 7$, this is a much more important effect since $\mathrm{FOV}_{\mathrm{interf}} \sim 15 \text{ mas}$ for $\lambda=1.7\text{ $\mu$m}$ and $B=100\text{ m}$; however, since we will be using the PIONIER data to explore the very close environment ($\lesssim 5 \text{ mas}$) of the A star, we can neglect this effect as well. In practice, this means that the PIONIER observations are completely insensitive to component B. 
\end{enumerate}

\subsection{VLTI/PIONIER}

As detailed above, due to the bandwidth smearing effect the PIONIER data is completely insensitive to the component B (the possibility that the projected separation of the star has changed significantly due to a high eccentricity and/or high inclination -- and therefore could be within the interferometric FOV of PIONIER -- does exist, but it will be shown below that that is not the case). On the other hand, the radial velocity variations of the A star suggest a period on the order of days, corresponding to a binary separation on the order of 0.1 AU $\leftrightarrow 1.3 \text{ mas}$. Therefore, we expect that the PIONIER observations with an angular resolution of 2.5 mas could be able to spatially resolve component A into a close binary. Indeed, upon inspection the calibrated interferometric data immediately showed a resolved source with squared visibilities below one and closure phases different than zero. We therefore proceeded to fit the binary model in Eq. \ref{eqn:binary} to the data, with $f_c=1$ since it is a very close binary and the fiber attenuation is therefore negligible. 

Because our interferometric data consists of squared visibilities and closure phases, the $\chi^2$ map produced by fitting Eqn. \ref{eqn:binary} to the data is not convex and has many local minima. In order to find the global minimum, we run 2d model grids in binary separation for a few different flux ratio to find the approximate region of the global minimum, and proceed with a gradient-based non-linear least squares minimization starting from the best solution in the grid to find the actual global minimum. This is standard practice in fitting a binary model to optical/NIR interferometric data \citep[see e.g.][]{Gallene15}. The fits are performed using the \texttt{python} package \texttt{lmfit}.

With the errorbars reported by the pipeline, the combined reduced $\chi^2$ of the best fit is quite high ($\chi^2_{\mathrm{red}} = 5.8$) and dominated by the squared visibilities; most likely, this is due to systematic errors in the calibration of the squared visibilities. On the other hand, we find very good agreement between data and model for the closure phases (which are more robust to calibration errors). We therefore increased the errors in the visibilities by a factor of 2, which produced a $\chi^2_{\mathrm{red}} = 1.9$ and more balanced residuals between the squared visibilities and closure phases. 

\begin{figure*}
 \includegraphics[width=2.2\columnwidth]{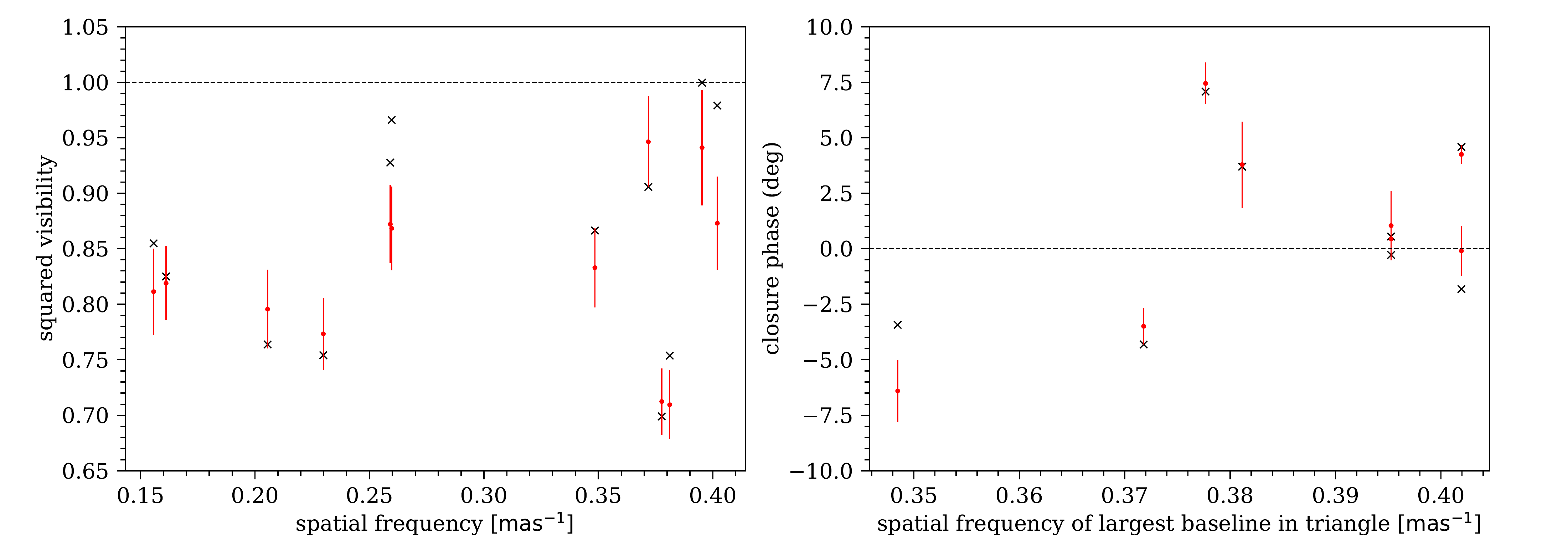}
 \caption{Data (red) and best fit binary model (black crosses) for the PIONIER observation of HIP87813. The dashed lines show the expected values for a single star, which is clearly ruled out.}
 \label{fig:PIONIER_fit}
\end{figure*}

Figure \ref{fig:PIONIER_fit} shows the PIONIER interferometric data and the best fit binary model, which has a flux ratio $f = 0.10 \pm 0.02$, separation $\rho = 1.7 \pm 0.2 \text{ mas}$ ($0.13 \pm 0.02 \text{ AU}$) and position angle PA$=318.2 \pm 0.9^{\degr}$. These uncertainties are the 1$\sigma$ values already scaled by $\sqrt{\chi^2_{\mathrm{red}}}$. From now on, we will refer to the primary star (i.e. the A-type star) in component A as A$_1$ and the secondary as A$_2$. 

\subsection{VLTI/GRAVITY}

\subsubsection{Acquisition camera image} 

Although the interferometric GRAVITY data were obtained only for the A star, the acquisition camera of the instrument (which operates in the NIR H band) also shows component C. In Figure \ref{fig:NACO} we plot the acquisition camera image averaged over the four telescopes in the bottom left. From the image we measured the separation between components (A+B) and C, which is reported in Table \ref{table:C_separation}, and a magnitude contrast $\Delta m_H = 7.00 \pm 0.05$. 

The module NAOMI provides AO correction which under ideal conditions returns a PSF close to the diffraction limit for the 1.8 m ATs in the H band (FWHM around 250 mas). However, component A is so bright that due to saturation the effective PSF in the acquisition camera image is much broader (FWHM around 1"). In the bottom right panel of Figure \ref{fig:NACO}, we show the residual image after subtracting the best-fit elliptical Moffat profile to the image. Unfortunately, component B cannot be resolved within the residuals of the very broad saturated PSF. 

\subsubsection{Interferometric data} 

\begin{figure*}
 \includegraphics[width=2\columnwidth]{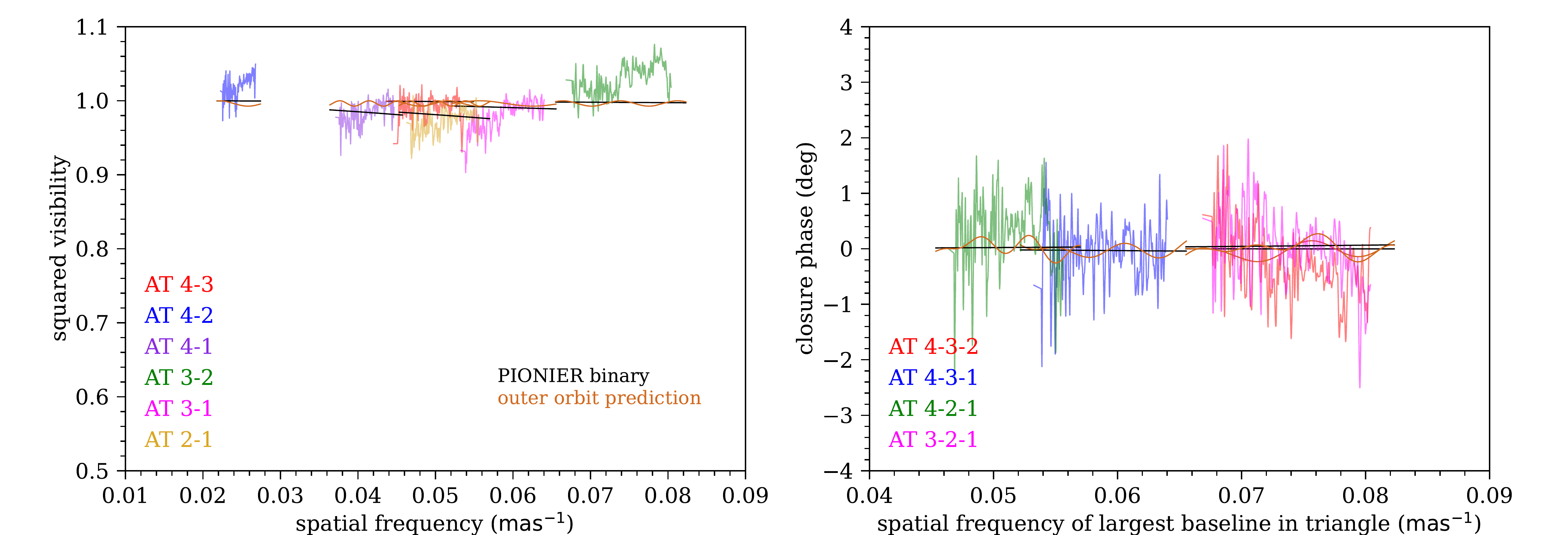}
 \caption{Science channel interferometric data for the GRAVITY observation of HIP 87813, with each baseline/triangle shown in a different color. The black and orange lines show the interferometric signatures created by the best-fit model for the inner binary A$_1$+A$_2$ detected in PIONIER and for the binary A+B predicted from the outer orbit in Section 5.}
 \label{fig:GRAVITY_vis}
\end{figure*}

The GRAVITY data consists of squared visibilities very close to unity and closure phases very close to zero (i.e. consistent with a single star), and do not immediately reveal the presence of any additional companion. In Fig. \ref{fig:GRAVITY_vis} we plot the interferometric data (squared visibilities and closure phases) for the SC channel for one of the GRAVITY files. The fact that the very close companion detected in the PIONIER observation is not detected in the GRAVITY data is not surprising due to the lower spatial resolution of the latter. We show in black in Figure \ref{fig:GRAVITY_vis} the interferometric signatures created by the best-fit binary model to the PIONIER data: clearly, the signatures are too small to be distinguished from a single star model within the errors. 

Contrary to the PIONIER data, however, the interferometric FOV of the GRAVITY data is very large (30"), and therefore it is sensitive to wide companions out to a separation of about 300 mas. This separation limit is imposed by the fiber coupling factor: a source located at $150$ mas from the center of light will be attenuated by about $50 \%$ and one at $300$ mas by about $94 \%$. Therefore, it was somewhat of a surprise that the component B seen in the NACO images was not seen in the GRAVITY interferometric data. This implies that the separation between components A and B was larger in the GRAVITY epoch so that the flux from B got significantly attenuated. 

\subsubsection{Companion detection limits} 

\begin{figure}
 \includegraphics[width=\columnwidth]{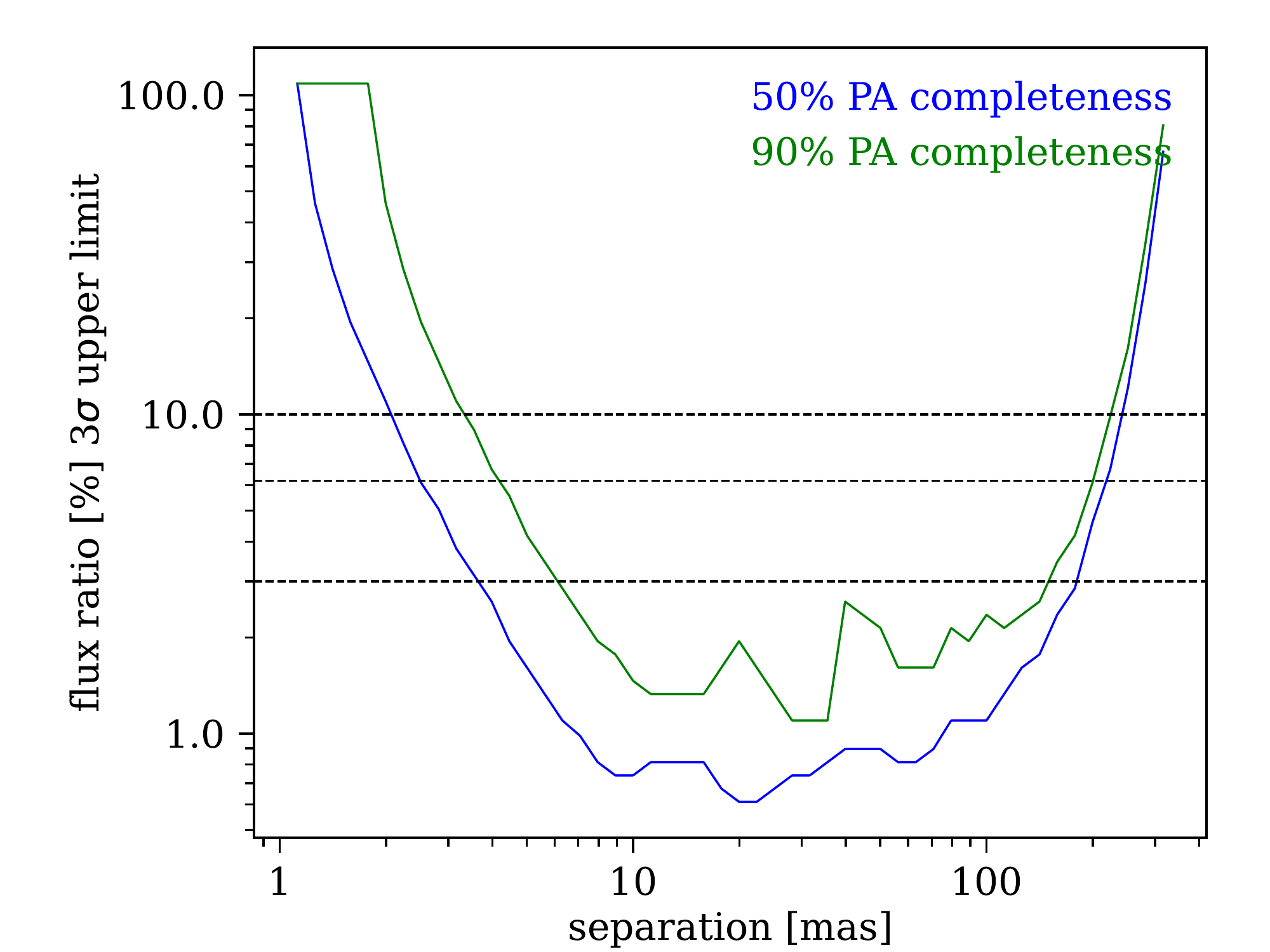}
 \caption{Companion detection limit for the GRAVITY data as a function of separation. The blue and green curves show the maximum flux ratio allowed for 50\% and 90\% of the PAs respectively. The horizontal lines demark flux ratio upper limits of 3, 6.2 and 10\%.}
 \label{fig:GRAVITY_flux_limit}
\end{figure}

We follow \cite{Absil11} to calculate flux ratio upper limits for a companion as a function of separation. Namely, we first re-scale the error bars so that $\chi^2_{red} = 1$ for a single star model. Next, we construct a 2D grid in $1 \text{ mas} < \rho < 320 \text{ mas}$ and $0^{\degr} < \text{PA} < 360^{\degr}$ and, for each point on the grid, inject an additional source in the model and find the flux ratio at which the model becomes inconsistent with the data at the $3\sigma$ level. Due to the sparse $uv$-coverage, the flux ratio upper limit at a given separation depends on the PA (in other words, there are specific positions where the source can be injected and lead to smaller interferometric signatures). Therefore, for each separation we show the median and the $90\%$ flux limits over the full range of PAs. Fig. \ref{fig:GRAVITY_flux_limit} shows the resulting companion detection limits from the GRAVITY data. 

In general, the contrast upper limit degrades at low separations due to the limited angular resolution of the array and at large separations due to the attenuation of the light from the source coupled into the GRAVITY fiber. The results in Fig. \ref{fig:GRAVITY_flux_limit} corroborate the fact that the companion with flux ratio 10\% at separation $\rho = 1.7 \text{ mas}$ detected in the PIONIER observations cannot be seen in the GRAVITY data. It also shows that for component B (detected in NACO images) with a flux ratio 6.2\% not to be seen in the GRAVITY data, its separation must be $\rho \gtrsim 200 \text{ mas}$. Using the 90\% completeness curve, the GRAVITY data also allows to exclude any additional companion with a flux ratio $f > 3\%$ within $7 \text{ mas} \lesssim \rho \lesssim 150 \text{ mas}$. 

\section{Putting it all together: the components of HIP 87813}

\subsection{The very faint companion is a background source} 

\begin{figure}
 \includegraphics[width=\columnwidth]{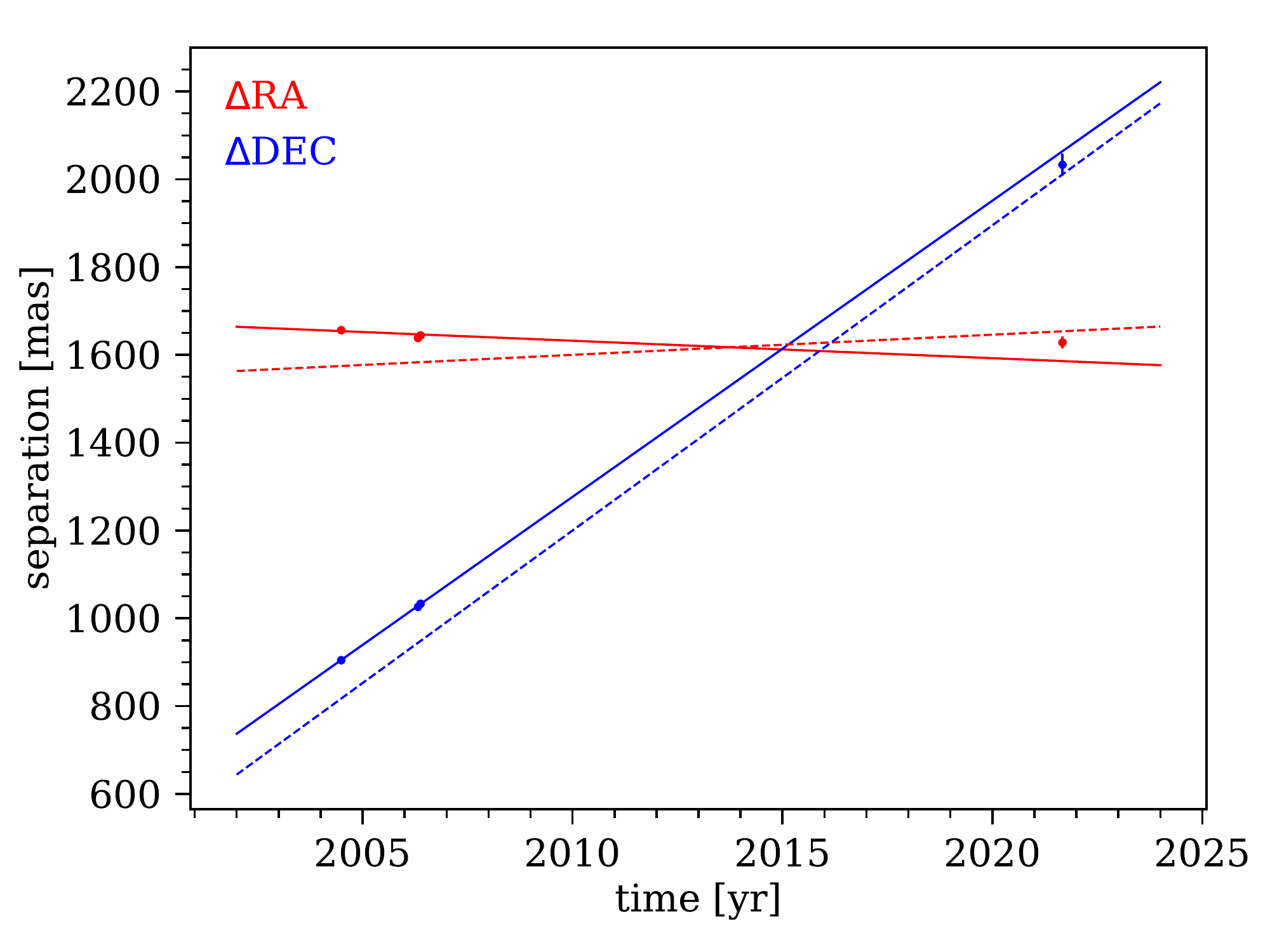}
 \caption{Separation between components A and C measured from the three NACO images and between A+B and C from the GRAVITY acquisition camera image as a function of time. The full lines show the best-fit linear motion, while the dashed lines show the opposite of the proper motion of HIP 87813 (with an arbitrary vertical shift).}
 \label{fig:M_star}
\end{figure}

We plot the separation between components C and A(+B, in the case of the GRAVITY image) as a function of time in Figure \ref{fig:M_star}. The data is well-fit by a linear motion of $-4.0 \pm 1.9 \text{ mas} \text{ yr}^{-1}$ in RA and $+67.5 \pm 0.8 \text{ mas} \text{ yr}^{-1}$ in DEC over 18 years. We note that a small systematic error in the GRAVITY separation is expected due to the fact that components A and B cannot be resolved and due to saturation of the camera. 

This motion is quite close to the opposite of the proper motion of HIP 87813 (which we find precisely in Section \ref{section:outer_orbit} below). Furthermore, it is not consistent with a Keplerian orbit: the characteristic velocity of the (supposedly) M star around the A+B system would be $v \sim \frac{2 \pi 200 \text{AU}}{1450 \text{ yrs}} \sim 4 \text{ km} { s}^{-1} $, where an orbit size of 2.6" $\rightarrow 200 \text{ AU}$ and the corresponding period for a total mass $\sim 3.8 M_{\odot}$ were used, which is about a factor of seven smaller than the linear motion measured. Therefore, we conclude with certainty that component C is not physically related to HIP 87813, and is instead a background source with a much smaller proper motion. This is not entirely surprising given that the sky region is quite crowded. Using the 2MASS catalogue \citep{Skrutskie06} we calculated that the number density of sources with $K<13(14)$ within 5' of HIP 87813 is $1.7(3.1) \times 10^{-3} \text{ arcsec}^{-2}$, which correspond to a non-negligible number of about 0.02(0.04) background sources in a 2" circle around it. 

\subsection{The masses of the components}

From the data presented above, it is clear that HIP 87813 is a hierarchical triple system consisting of an inner binary (A = A$_1$ + A$_2$) and an outer tertiary (B). From the PIONIER data we found $\frac{f_{A_2}}{f_{A_1}} = 0.10 \pm 0.02$ in the H band (the K band flux ratio is the same within the errors; see below) and from the NACO images we found $\frac{f_{B}}{f_{A_1}+f_{A_2}} = 0.062 \pm 0.002$ in K band. From the total apparent K band magnitude in 2MASS $K=5.70$ and a distance $d=78.2 \text{ pc}$, the total absolute magnitude is $K=1.24$. We can therefore solve for the absolute K band magnitude of each component. We find $K_{A_1} = 1.41 \pm 0.02$, $K_{A_2} = 3.9 \pm 0.2$ and $K_{B} = 4.33 \pm 0.04$.

For stars with $2.5 \lesssim M_K \lesssim 5.5$ ($0.6 M_{\odot} \lesssim M \lesssim 1.5 M_{\odot}$), there exists a rather tight log-linear K band magnitude-mass relation \citep[e.g.][]{Henry93}, with small deviations for very young stars (age $\lesssim 100 \text{ Myrs}$). Using Figure A.5 (right) in \cite{DeRosa14}, which was constructed from theoretical isochrones from \cite{Baraffe98}, we find an approximate relation 

\begin{align}
\log M [M_{\odot}] \approx -0.145 M_{K} + 0.496  
\end{align}

\noindent Applying this to stars A$_2$ and B, we find masses $M_{A_2} = 0.85 \pm 0.06$ (G8V-K2V) and $M_B = 0.74 \pm 0.01$ (K3V). Note that the PIONIER flux ratio is in the H band and not in the K band. Approximating star A$_1$ as a blackbody with $T \sim 9290 \text{ K}$ and star A$_2$ with $T \sim 5240 \text{ K}$, the ratio between the flux ratio in the H band to that in the K band is $90\%$. Such a $10\%$ correction is only half of the flux ratio error, and therefore negligible in this case. 

For A stars, the age effect is much more important and a precise mass cannot be obtained from the absolute K band magnitude alone. From Figure A.5 (left) in \cite{DeRosa14}, the mass of A$_1$ for its $M_K=1.41$ is in the range $1.8-2.7 M_{\odot}$ for ages between 63 Myrs and 1 Gyr. Based on the $M_{V_T}-M_K$ color (where $V_T$ is the Tycho visual magnitude) and using theoretical isochrones from \cite{Siess00} (for solar metallicity and a non-rotating star), \cite{DeRosa14} estimated a mass $M_{A_1} \sim 2.11 M_{\odot}$ and age $t=350 \text{ Myrs}$ for the A star. However, the two components A$_2$ and B make the A star appear redder (and therefore older) than it really is. Using an approximate relation between $M_{V_T}-M_K$ and $M_K$ derived from close main sequence stars with $M_K \sim 4$ in $Gaia$, 2MASS and Tycho:

\begin{align}
M_{V_T} - M_{K} \approx 1.3 M_{K} - 2.8 
\end{align}

\noindent we estimate $M_{V_T}$ for stars A$_2$ and B, and find the real $M_{V_T} \approx 1.49$ for the A star alone using the total $M_{V_T}=1.47$ (as expected, the change in $M_{V_T}$ is much smaller than $M_K$ because the contaminants are lower mass stars). Finally, using the corrected $M_K=1.41$ and $M_{V_T}-M_K=0.08$ for the A star we find an age $log(t) \approx 8.25 \Rightarrow t \approx 180 \text{ Myrs}$ using Fig. A.3 in \cite{DeRosa14}, and a corresponding mass $M_{A_1} \approx 2.2 M_{\odot}$ using their Fig. A.4. The isochrones ignore rotation so that, although the A star is not rotating very fast ($v \sin i \approx 60 \text{ km} \text{ s}^{-1}$), this age estimate should be taken as an upper limit. Furthermore, because this mass estimate is not free of systematic errors, we do not fix the mass of the A star in the further analysis, but a lower limit $M_{A_1} > 1.8 M_{\odot}$ is quite robust.

\subsection{The RV curve of the inner binary}
\label{section:RV}

The A star is clearly variable in radial velocity, pointing to a binary system with a period on the order of days. We fit the RV data with a model for a Keplerian orbit: 

\begin{equation}
v_{z,1} = - K_1 (e \cos (\omega + \pi) + \cos ( (\omega+\pi) + \nu) ) + \gamma 
\end{equation}

\noindent where $K_1$ is the radial velocity semi-amplitude of the primary star (A$_1$), $e$ is the eccentricity, $\omega$ is the argument of periastron, $\nu$ is the eccentric anomaly (which involves the other two parameters of the solution, orbital period $P_{\mathrm{orb}}$ and time of periastron $T_p$) and $\gamma$ is the systemic velocity. Note that we define $\omega = \omega_2$ as the argument of periastron of the secondary star (A$_2$), so that $\omega_1 = \omega + \pi$. This consistency is important when combining the RV and interferometric data below. The best solution is found assuming Gaussian statistics for the errors through $\chi^2$ minimization using gradient-based nonlinear least squares using the \texttt{python} package \texttt{lmfit}. In order to find the global minimum, we perform fits over a finely sampled 2d grid in $P_{\mathrm{orb}}$ and $e$, estimating initial values for the other parameters at each grid point using the data \citep[e.g.][]{Milson20}. The latter is constraining enough that a clear global minimum exists for the orbital solution. Although the errorbars for the literature and especially for the FEROS RVs are much smaller than the SpeX RVs, the latter have a crucial role in the fit because they provide orbital phase coverage that would otherwise be missing. 

Furthermore, we also allowed for an absolute velocity shift between the different data sets (the three \cite{Nordstrom85} points, the \cite{Grenier99} point, the FEROS points and the SpeX and GRAVITY points) in order to take into account possible systematic errors in the zero points between the different sets, caused for e.g. by instrumental absolute wavelength calibration errors or errors in the effective rest wavelength of the different absorption lines used (intrinsic or due to blending of weaker lines). Shifts could also be caused in the presence of a longer period orbit, since the different data sets are acquired around a decade from each other. The best-fit shifts of the datasets relative to the FEROS dataset are reported in Table \ref{table:RV} and are consistent with or very close to zero. 

\begin{figure}
 \includegraphics[width=\columnwidth]{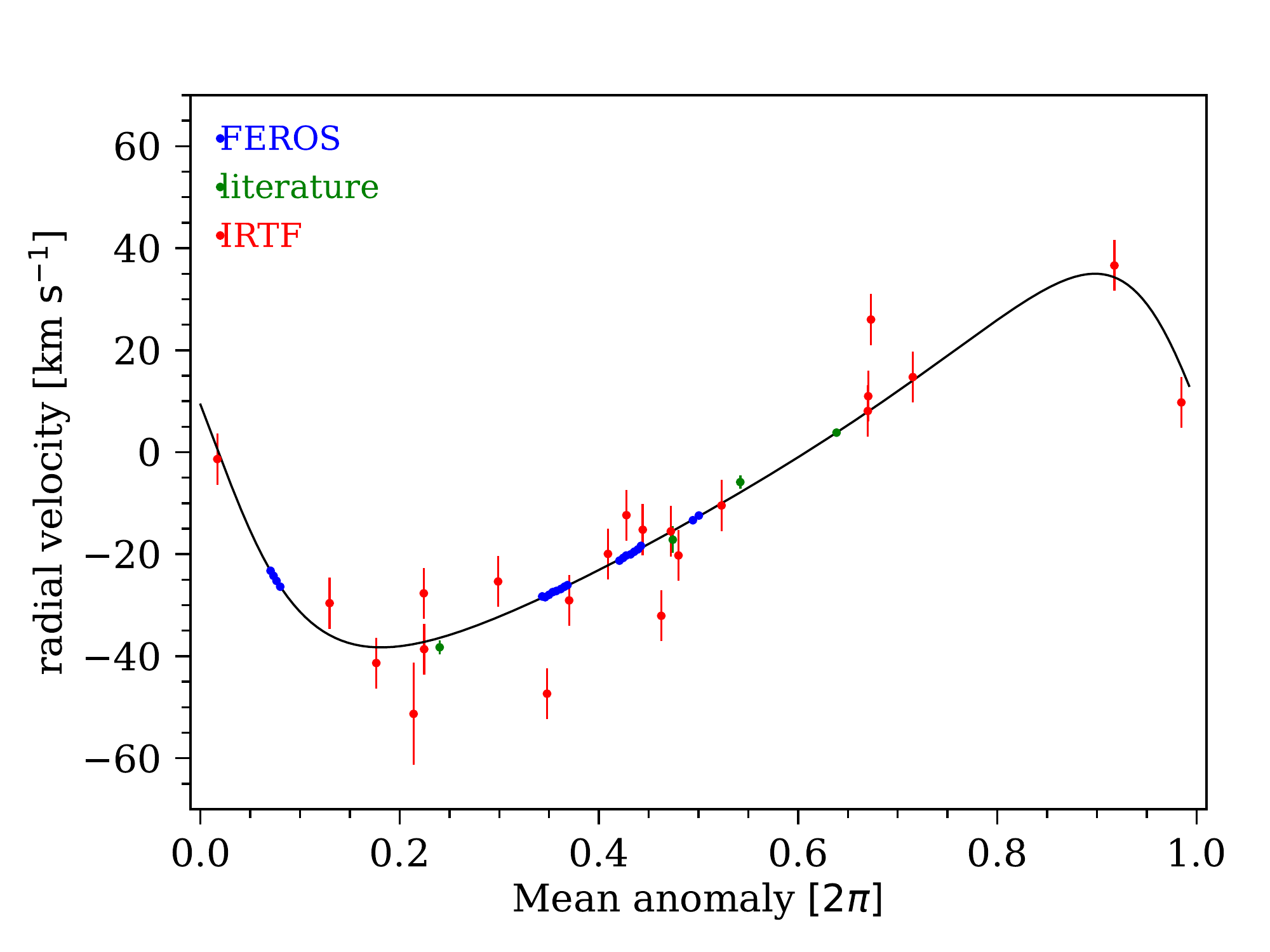}
 \caption{Data and best-fit radial velocity curve for the orbit of the A-type star A$_1$ within the close binary A.}
 \label{fig:RV}
\end{figure}

The reduced $\chi^2$ of the best-fit RV curve is $4.5$, which suggests the errorbars may be underestimated by around a factor of two. The best-fit parameters and errors (already scaled by $\sqrt{\chi^2_{red}}$) are reported in Table \ref{table:RV}, and the best fit is shown in Fig. \ref{fig:RV} (with the absolute shifts subtracted). 

\begin{table*}[t]
\centering
\caption{\label{table:RV} Best-fit radial velocity curve parameters for the A$_1$+A$_2$ binary.}
\begin{tabular}{cc}
\hline \hline
$K_1$ & $36.6 \pm 3.4 \text{ km}\text{ s}^{-1}$ \\[0.3cm]
$e$ & $0.36 \pm 0.03$ \\[0.3cm]
$P_{\mathrm{orb}}$ & $13.4193 \pm 0.0006 \text{ days}$ \\[0.3cm]
$\omega$ & $72.6 \pm 5.5^{\degr}$ \\[0.3cm]
$T_p$ & JD $2457900.1 \pm 0.2$ \\[0.3cm]
$\gamma$ & $-5.6 \pm 1.7 \text{ km}\text{ s}^{-1}$ \\[0.3cm]
\hline
$\Delta_{vz}$ \cite{Nordstrom85} & $-3.8 \pm 3.5 \text{ km}\text{ s}^{-1}$ \\[0.3cm]
$\Delta_{vz}$ \cite{Grenier99} & $0.1 \pm 2.0 \text{ km}\text{ s}^{-1}$ \\[0.3cm]
$\Delta_{vz}$ SpeX/GRAVITY & $-5.5 \pm 2.6 \text{ km}\text{ s}^{-1}$ \\[0.3cm]
\hline 
\end{tabular}
\tablefoot{}
\end{table*}

We find that the orbit is moderately eccentric with $e=0.36\pm0.03$ and a period $P=13.4 \text{ days}$. The resulting mass function is 

\begin{align}
\frac{M_{A_2}^3}{(M_{A_2}+M_{A_1})^2} (\sin i)^3 = \frac{P_{\mathrm{orb}}}{2 \pi G} K_1^3 (1-e^2)^{3/2} \\ = 0.055 \pm 0.016 M_{\odot}
\end{align}

\noindent Using $M_{A_2} = 0.85 \pm 0.06 M_{\odot}$, the semi-major axis is $a = 0.153 - 0.168 \text{ AU}$ for $M_{A_1} = 1.8 - 2.7$. For the minimum A$_1$ mass of $1.8 M_{\odot}$, we have $i = 54^o \pm 10^o$ (or $i=126^o \pm 10^o$). For $M_{A_1} = 2.2 M_{\odot}$ as estimated from the theoretical isochrones, we have $i \approx 70^o \pm 20^o$ or $110^o \pm 20^o$. The minimum inclination for the binary to be eclipsing is about $i \gtrsim \arccot \left ( \frac{R_{A_1}+R_{A_2}}{a} \right ) \approx \arccot \left ( \frac{1.9 R_{\odot} + 0.9 R_{\odot}}{0.16 \text{ AU}} \right ) \approx 85^o$. While we cannot exclude that A$_1$+A$_2$ is an eclipsing binary, it is probably not. 

The projected semi-major axis of the A star around the center of mass of the binary 

\begin{equation}
a_1 \sin i = K_1 \frac{P}{2 \pi} \sqrt{1-e^2} \approx 0.045 \text{ AU} \leftrightarrow 0.6 \text{ mas} 
\end{equation}

\noindent matches the astrometric noise (0.62 mas) of HIP 87813 in $Gaia$ eDR3, and is therefore very likely its dominant cause. The size of the semi-major axis on sky is $a \approx 0.16 \text{ AU} \leftrightarrow 2 \text{ mas}$ and can be readily identified with the binary resolved by PIONIER. 

\subsection{The orientation of the inner binary} 

Although a single astrometric point is not enough to completely determine the remaining two Keplerian parameters of the A$_1$+A$_2$ orbit (i.e. the inclination $i$ and the longitude of the ascending node $\Omega$), it can be used to significantly constrain them (and therefore the 3d orientation of the binary). We do this by building a grid in $i \in [0^0,180^o]$ and $\Omega \in [0^o,360^o]$ and finding the difference between the predicted (from the Keplerian orbit) and the measured $\Delta\text{RA} = -1.13 \pm 0.13 \text{ mas}$ and $\Delta\text{DEC} = 1.27 \pm 0.15 \text{ mas}$ in the PIONIER observation. The remaining orbital parameters are taken from the best-fit model to the RV curve, and $M_{A_2} = 0.85 M_{\odot}$ is also used (so that $M_{A_1}$ is calculated from the mass function and the inclination). We note that the resulting uncertainty in $i$ and $\Omega$ is much larger than for the other orbital parameters determined from the RV curve, so that fixing the latter to their best-fit values (rather than performing a joint fit) is appropriate.  

\begin{figure*}
 \includegraphics[width=2.2\columnwidth]{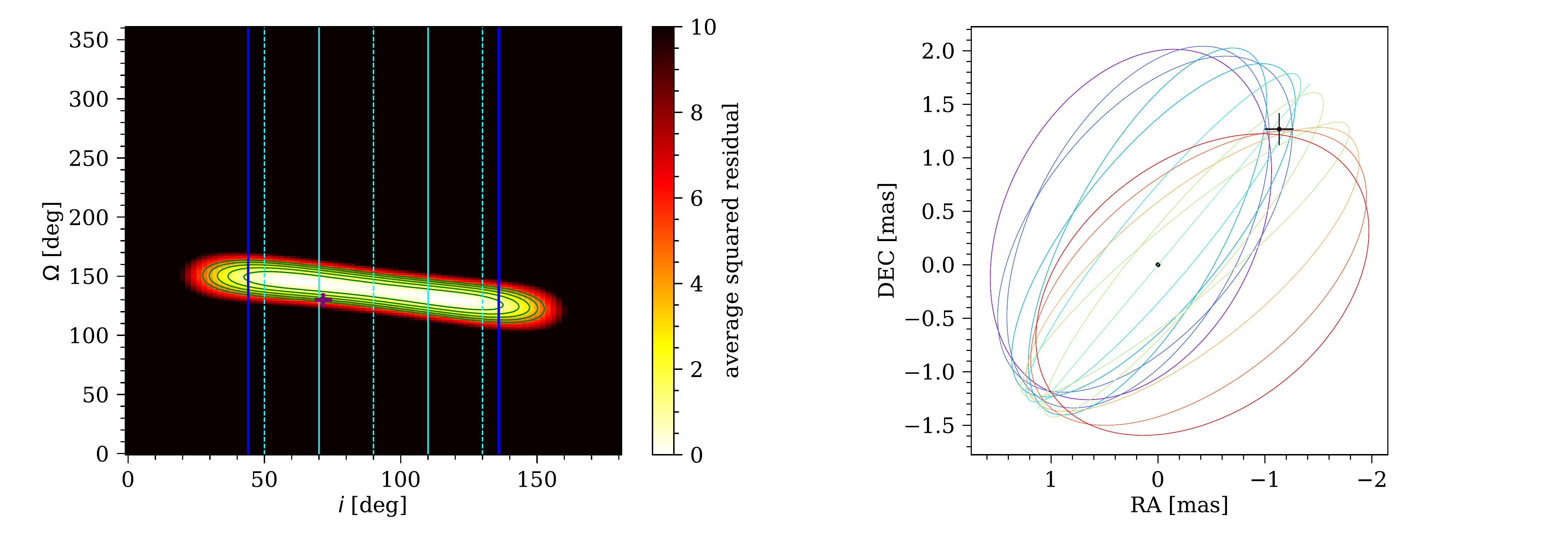}
 \caption{\textbf{Left}: Average squared residual for the on-sky separation of the A$_1$+A$_2$ binary in the PIONIER observation as a function of the inclination and longitude of the ascending node. The green lines mark the 1, 2, 3, 4 and 5 contours (for the blue vertical lines see text). The magenta point marks the orientation of the outer orbit A+B. \textbf{Right}: Locus of  a few possible projected binary orbits that give an average squared residual to the PIONIER point (black) less than one, for inclinations varying from $50^o$ to $130^o$ in rainbow order. The primary star A$_1$ is at (0,0).}
 \label{fig:i_Omega_grid}
\end{figure*}

In Figure \ref{fig:i_Omega_grid} (left) we plot the resulting averaged (over $\Delta\text{RA}$ and $\Delta\text{DEC}$) squared residual $\chi^2_{\mathrm{avg}}$. The green lines mark the $\chi^2_{\mathrm{avg}}=$1, 2, 3, 4 and 5 contours. There is clearly a large correlation between $i$ and $\Omega$, and $\Omega$ is particularly well-constrained. Taking the $\chi^2_{\mathrm{avg}}=2$ contour, for example, the allowed range is $35^o \lesssim i \lesssim 145^o$ and $120^o \lesssim \Omega \lesssim 160^o$. The vertical blue lines mark the (1$\sigma$) inclination range allowed for a minimum A star mass of 1.8 $M_{\odot}$, while the vertical cyan lines mark the inclination (plus 1$\sigma$ ranges) for $M_{A_1} = 2.2 M_{\odot}$ as estimated from the theoretical isochrones.

In Figure \ref{fig:i_Omega_grid} (right), we plot a few projected orbits for which $\chi^2_{\mathrm{avg}}<1$, with the PIONIER measurement in black and the primary star A$_1$ at the origin, in order to illustrate the procedure. The inclinations vary from $50^o$ to $130^o$ in rainbow order. In the near future, either more interferometric observations or the $Gaia$ DR3 release will probably fully determine the orbit. 

\subsection{The orbit of the outer binary} 
\label{section:outer_orbit} 

In this section, we constrain the parameters of the outer orbit between A=A$_1$+A$_2$ and B, establishing HIP 87813 as a hierarchical triple system. We do this by combining the separations measured in the three NACO images with the proper motions listed in Table \ref{table:PM} (except for the FOCAT-S point since we do not know how apart the epochs used to derive the proper motion are separated) and the positions listed in Table \ref{table:positions}. For the latter, besides $Hipparcos$ and $Gaia$, we also use a position from the Guide Star Catalog (GSC) version 2.3.2 \citep{Lasker08} with 10 mas error. We take the $Hipparcos$ position as the reference position (only positional differences matter for the fit). The parameters consist of six orbital parameters, the mass of star A$_1$ and the two intrinsic linear proper motions in RA and DEC. Combining all the data, there are 12 measurements for the positions (RA and DEC), 8 measurements for the proper motions and 3 measurements for the direct separation. 

Because $Hipparcos$, $Gaia$ and GSC do not resolve components A and B, they measure the motion of the center of light (COL) around the center of mass (COM). It traces the same Keplerian motion as the primary star but with a semi-major axis somewhat reduced. The semi-major axis of the COL relative to the COM is 

\begin{align}
a_{\mathrm{COL}} = \left | \frac{-f_A a_A + f_B a_B}{f_A + f_B} \right | = a_A \frac{1 - \frac{f}{q}}{1+f}
\end{align}

\noindent where $f=\frac{f_B}{f_A}$ is the flux ratio and $q=\frac{M_B}{M_A}$ is the mass ratio between the components. For identical stars, there is no astrometric motion. In our case, with $f=0.062$ and $q\sim0.25$, we expect the COL to follow the primary "star" (A=A$_1$+A$_2$) with about a $20\%$ reduction in semi-major axis.

We fit for the Keplerian orbit that can best explain all the data. We use linear least squares implemented with the \texttt{python} package \texttt{lmfit} and run fits over a fine eccentricity grid between 0.01 and 0.99 (we find that keeping the eccentricity fixed helps to prevent the fit from diverging). We find a well-defined global minimum with $\chi^2_{\mathrm{red}} = 4.55$ and its parameters and errors (scaled by $\sqrt{\chi^2_{\mathrm{red}}}$) are reported in Table \ref{table:outer fit}. We find the outer orbit to be mildly eccentric ($e=0.26 \pm 0.03$) with a period $P_{\mathrm{orb}}=60.8 \pm 12.0 \text{ yrs}$. We also note that the mass of the A star, which was let free to vary, has a best fit value $M_{A_1} = 1.61 \pm 0.51 M_{\odot}$, consistent with the allowed range $M_{A_1} > 1.8 M_{\odot}$ but too uncertain for a useful constraint. The resulting semi-major axis of the outer orbit is $a \approx 24 \text{ AU}$.

\begin{table}[t]
\centering
\caption{\label{table:outer fit} Best-fit parameters for the outer orbit A+B.}
\begin{tabular}{cc}
\hline \hline
$e$ & $0.26 \pm 0.03$ \\[0.3cm]
$i$ & $71.6^o \pm 3.2^o$ \\[0.3cm]
$\omega$ & $160.5^o \pm 19.6^o$ \\[0.3cm]
$\Omega$ & $130.1^o \pm 5.5^o$ \\[0.3cm]
$P_{\mathrm{orb}}$ & $60.8 \pm 12.0 \text{ yrs}$ \\[0.3cm]
$T_p$ & $1997.2 \pm 2.1$ \\[0.3cm]
$M_{A_1}$ & $1.61 \pm 0.51 M_{\odot}$ \\[0.3cm]
pmra & $-4.6 \pm 0.4 \text{ mas} \text{ yr}^{-1}$ \\[0.3cm]
pmdec & $-69.5 \pm 0.9 \text{ mas} \text{ yr}^{-1}$ \\[0.3cm]
\hline 
\end{tabular}
\end{table}

\begin{figure*}
 \includegraphics[width=2\columnwidth]{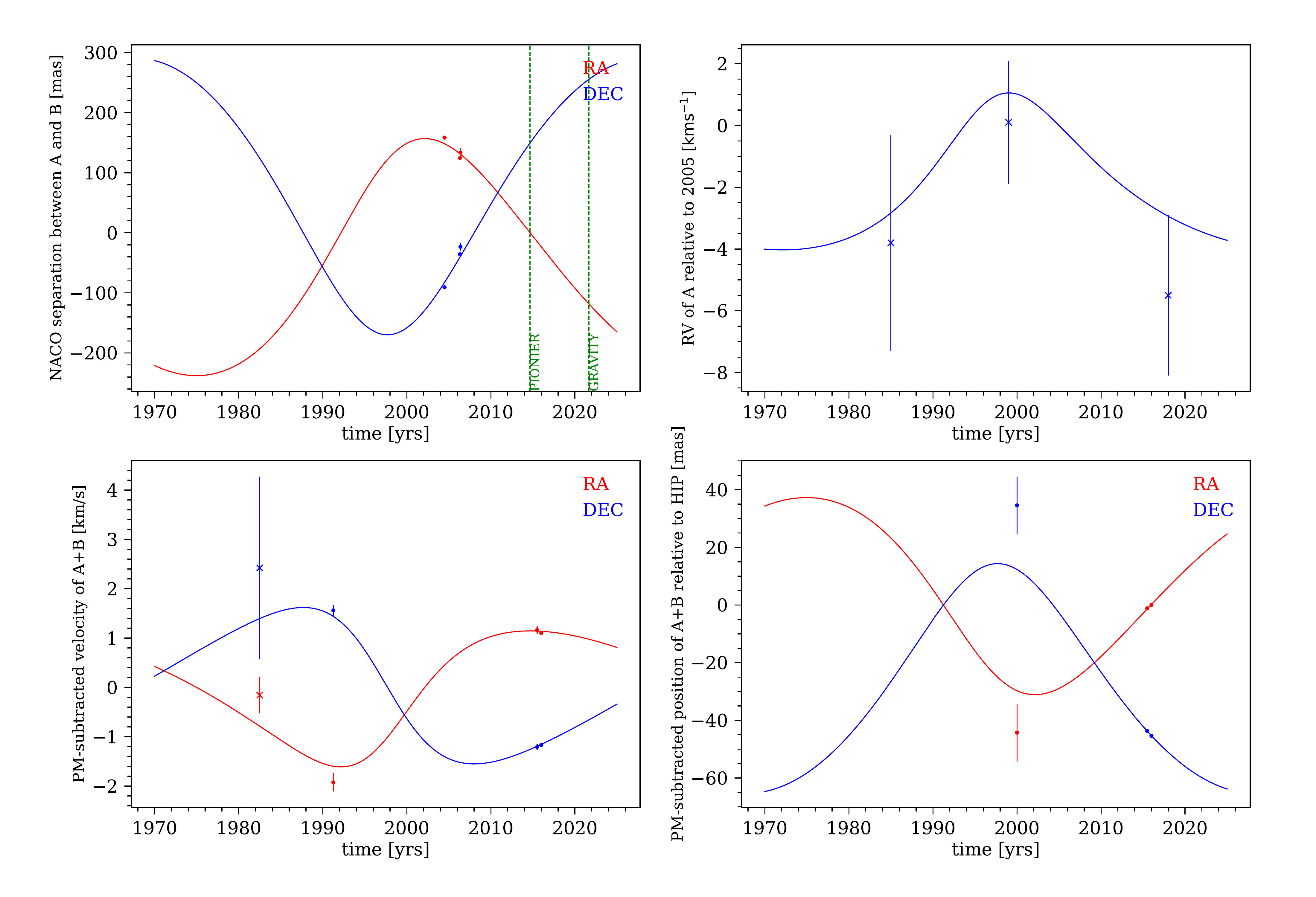}
 \caption{Data and best fit model for the outer orbit of component A=A$_1$+A$_2$ around B. \textbf{Upper left}: direct separation measured from NACO images. The vertical green lines mark the epochs of NIR interferometric PIONIER and GRAVITY observations. \textbf{Lower left}: tangential velocities (proper motion subtracted). \textbf{Lower right:} positions relative to $Hipparcos$ (proper motion subtracted). \textbf{Upper right:} radial velocity relative to 2005 (the FEROS epoch), which were not used for the fit.}
 \label{fig:outer_orbit}
\end{figure*}

In Figure \ref{fig:outer_orbit}, we plot the data and the best fit for the outer orbit. The upper left panel shows the separation from the NACO images, the bottom left panel shows the tangential velocities (already with the best-fit proper motion subtracted) and the bottom right panel shows the positions (relative to the $Hipparchos$ position and also subtracted for the best-fit proper motion), with the FOCAT-S points (not used in the fit) marked with "x". In the upper right panel, we show the velocity shifts measured from the RV curve (relative to the FEROS dataset in 2005) compared to the model. While we decided to not use these data in the fit for the reasons discussed in Section \ref{section:RV}, they are consistent with the model (within their large errors relative to the orbital motion). 

Finally, we calculate the predicted separation between components A and B at the time of the PIONIER and GRAVITY interferometric observations, shown as the vertical green lines in the upper right panel of Figure \ref{fig:outer_orbit}. The predicted separation at the PIONIER epoch is 150 mas, so component B was indeed well outside the interferometric FOV and invisible to PIONIER. Meanwhile, the predicted separation at the GRAVITY epoch is 282 mas. We plot the interferometric signature that the B component at this separation would create in the GRAVITY data in Figure \ref{fig:GRAVITY_vis}. Indeed, the increase in separation explains why B could not be detected in the GRAVITY data. 

\subsection{The relative orientation between the inner and outer orbits}

In Figure \ref{fig:i_Omega_grid} (left), we plot in magenta the orientation of the outer orbit measured in the previous section. Remarkably, it falls within the constraints of the orientation of the inner binary, and is consistent with the $\chi^2_{\mathrm{avg}}=2$ surface within $1\sigma$. It also falls very close to the inclination needed to make $M_{A_1} = 2.2 M_{\odot}$ as estimated from theoretical isochrones. This is strong albeit not definitive evidence that the inner and outer orbits are coplanar. 

In order to quantify this, we compute the volume of the sphere contained within the $\chi^2_{\mathrm{avg}}=2$ contour of the inner orbit orientation and find $4\%$. While there remain allowed regions for which the relative inclination is larger than $40^o$ (for e.g. if $i_{\mathrm{inner}} \sim 110^o$ and the orbits are ``mirrored''), we argue that the data is suggestive of coplanarity. A definitive answer should soon be reached once the orientation of the inner binary is more tightly constrained. 

\begin{comment} 
we show in Figure \ref{fig:relative orientation} the relative orientation $\phi_{\mathrm{rel}}$ between the inner and outer orbits as a function of $i_{\mathrm{in}}$ and $\Omega_{\mathrm{in}}$: 

\begin{align}
\cos \phi_{\mathrm{rel}} = \cos i_{\mathrm{in}} \cos i_{\mathrm{out}} + \sin i_{\mathrm{in}} \sin i_{\mathrm{out}} \cos ( \Omega_{\mathrm{in}} - \Omega_{\mathrm{out}} )
\end{align}

\noindent For this, we fix $i_{\mathrm{out}}$ and $\Omega_{\mathrm{out}}$ at their bet-fit value since their uncertainty is much smaller. While there remain viable regions for the inner orientation that result in a relative inclination $\phi_{\mathrm{rel}} > 40^o$ relative to the outer orbit, the fact that the orientation of the outer binary falls within the relatively narrow constraints of the inner binary (for example, the $\chi^2_{\mathrm{avg}}<2$ surface occupies  3.9\% of the sphere) is highly suggestive that they are closely aligned.
\end{comment} 

\section{Discussion}

We have shown that HIP 87813=HJ 2814A is a hierarchical triple. The inner binary (A=A1+A2) consists of a $\sim 2.2 M_{\odot}$ A-type star and a $0.85 M_{\odot}$ star on a very close ($a = 0.16 \text{ AU}$, $P=13.4 \text{ days}$) and midly eccentric ($e=0.36$) orbit. The outer companion (B) is a $0.74 M_{\odot}$ star on a $P \sim 60 \text{ yrs}$ ($a \sim 24 \text{ AU}$), also mildly eccentric ($e=0.26$) orbit. The system also orbits another solar-mass 2 AU binary, HJ 2814B (D=D1+D2), over a period $P\sim26 \text{ kyrs}$ ($a\sim 1600 \text{ AU}$). Finally, the faint star 2" away from HIP 87813 reported in \cite{DeRosa14} turned out to be a background source. A schematic summary of the system is shown in Figure \ref{fig:schematic}. The quintuple HJ2814 therefore joins a small list of high multiplicity systems containing an intermediate mass (A type) primary \citep[e.g.][]{Tokovinin97,Eggleton08,Tokovinin14,Tokovinin18}. 

This system is particularly interesting as it shows separations on three different length scales (triple hierarchy) and involves 5 stars that are sufficiently massive to evolve into WDs on the scale of a Hubble time.
%\citep[for a review on close binary formation, see e.g.][]{Kratter11}. 
%\subsection{Uniqueness of the system} \boaz I don't think we need a header for this issue just write the text here
It is interesting to compare it with some other 5+ multiples with A star primaries known. In the VAST survey \citep{DeRosa14}, limited to $d \lesssim 75 \text{ pc}$, there are only two quintuples reported (and no 6+ systems), HIP 11569 and HIP 90156 (both in the North and therefore not part of our VLTI survey). HIP 11569 (45 pc) contains one A star and four $\sim 1 M_{\odot}$ stars, but does not have a very close binary like HIP 87813 (the shortest period is 48.7 yrs). HIP 90156 (58 pc) has one A star and 2 $\sim 1 M_{\odot}$ stars but its quintuple status is not confirmed \citep[it is currently listed as a quadruple in the Updated Multiple Star Catalog;][]{Tokovinin18}. It is important to notice that the common proper motion search in \cite{DeRosa14} has been superseded by $Gaia$, and as a result many systems in the VAST survey have had their multiplicity upgraded with very wide companions in the Updated Multiple Star Catalog. 

Therefore, we also searched the latter for 5+ systems with A star primaries and found 2 septuples, 3 sextuples and about 10 quintuples that are at least reasonably well characterized. The majority is located within 200 pc of the Sun but three of them are beyond it. Their properties are varied but many of them show a large number of A or $\sim 1 M_{\odot}$ stars within a triple hierarchy structure (periods of a few to tens of days, decades to centuries, and few to tens of kyrs), for e.g. HIP 36850, HIP 21402, HIP 56109 and HIP 41564 within 100 pc, and HIP 107162 and HIP 58112 beyond it. Therefore, while HJ 2814 is rare it is certainly not unique, and these systems may share a common origin or fate. As far as we can tell none of these systems have been observed with interferometry, so that the strength of Kozai-Lidov oscillations within the inner hierarchy (see below) is not constrained. Quantitative estimates of the abundance and statistical properties of similar systems are needed to make progress but are beyond the scope of this paper.

\subsection{Dynamical processes in the quintuple system}
%- High hierarchy implies that the system is stable (caveat of high eccentricity of the outer orbit)
The constraints on the configuration obtained from the observations described above allow us to narrow down the dynamical behaviour of this system. 

\begin{figure}
 \includegraphics[width=\columnwidth]{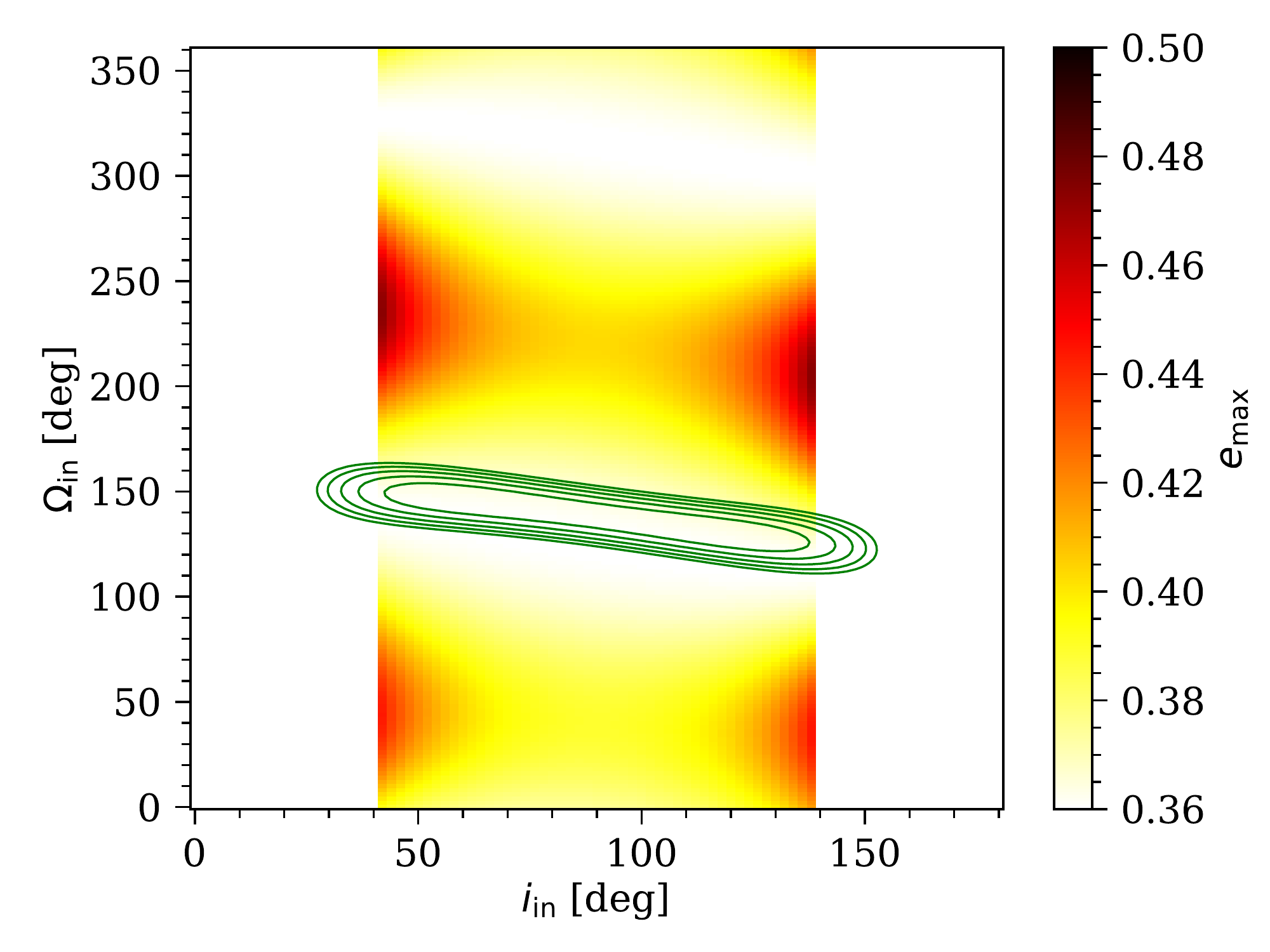}
 \caption{The maximum eccentricity attainable in A$_1$+A$_2$ through Kozai-Lidov oscillations induced by B (under the quadrupolar, averaged approximation). The green contours show the 1-5$\sigma$ constraints on the inner orbit, and inclinations for which $M_{A_1}<1.8 M_{\odot}$ have been excluded.}
 \label{fig:e_max_contour}
\end{figure}

The high hierarchy implies that the system is dynamically stable. Over short time scales, the 5 stars will move along the 4 Keplerian orbits involved (A1-A2,A-B,D1-D2,AB-D). Over long time scales, hierarchical systems can exchange significant angular momentum leading to precession and oscillations in eccentricity and inclination of the inner orbits (Kozai-Lidov oscillations \citet{Lidov62,Kozai62}, see \citet{Naoz16,Ito19} for recent reviews). In particular, excitation of the eccentricity by a third object has been argued to lead to orbital shrinkage due to tidal dissipation \citep[e.g.][]{Kiseleva98,Fabrycky07,Naoz14} and it is natural to ask if the close binary A1-A2 is currently experiencing such a process. The Kozai-Lidov oscilation time of the A1-A2 + B triple system is roughly given by $\tau_{KL}\sim P_{\mathrm{out}}^2/P_{\mathrm{in}}\sim 10^5$ years, where $P_{\mathrm{out}}=60$ years is the outer A-B orbital time and $P_{\mathrm{in}}=13.4$ days is the inner A1-A2 orbital time, and is certainly short enough. In order that eccentricity changes significantly, the mutual inclination has to be sufficiently high and the oscillations need to be faster than the General Relativistic (GR) precession time.  For the masses and separation scales of this triple system, a hierarchy $a_{\mathrm{out}}/a_{\mathrm{in}}$ smaller than $\sim 100$ is required to overcome GR precession \citep[using Eq. 5 in][with low outer eccentricity]{Dong14}. The measured hierarchy of 24AU/0.16AU=150 is marginal implying that oscillations may be possible depending on the mutual inclination and requiring a detailed determination of the orbital parameters. Despite the significant degeneracy in the orientation of the A1-A2 orbit (Figure \ref{fig:i_Omega_grid}) a tight constraint on the eccentricity oscillations can be obtained as shown in Figure \ref{fig:e_max_contour} and described next. 

For each choice of orientation of the A1-A2 orbital plane ($i_{\mathrm{in}},\Omega_{\mathrm{in}}$), the parameters of the triple system A1-A2+B are completely set and the maximal eccentricity that can be achieved in the Kozai-Lidov oscillation can be algebraically found in the quadruple approximation (which is excellent given the high hierarchy). This is obtained by requiring that the sum of the Quadruple and GR Hamiltonians is conserved \citep[e.g. Eqs. 12,22 in][]{Fabrycky07}, as well as the outer eccentricity \citep{LidovZiglin76} and the total angular momentum, and finding the maximal eccentricity among the extrema at $\omega_{\mathrm{in}}=\pi/2$. As can be seen in figure \ref{fig:e_max_contour}, the oscillations obtained within the allowed orientations are negligible. At the present eccentricity, it is likely that the A1-A2 system is experiencing negligible tidal dissipation within a timescale of the age of the system ($\sim 200$ Myr): tidal dissipation in A2 is unlikely as the circularization period of the larger sun-like stars over billions of years is likely shorter than 13 days \citep[e.g.][]{Zahn77,DuquennoyMayor91,Mathieu04,Raghavan10,Zanazzi21} and is a strong function of the period and radius of the star; tidal migration due to dissipation in the primary A1 A-star is largely ruled out by its fast rotation ($v_{\mathrm{rot}} \sin i = 52 \text{ km}\text{ s}^{-1}$ ) with a rotation period $p\lesssim 2$days which is much shorter than the synchronization period \citep[and the pesudo-syncronization period;][]{Hut82} - if tidal dissipation was significant, synchronization would have been achieved on much shorter time scales than circularization. The low tidal dissipation in the A-star and lack of synchronization is compatible with estimates for the parameters of the system for a radiative envelope \citep[][]{Zahn77}. 

The behaviour of the A1-A2+B system may be significantly affected by the D system if the pericenter of the AB-D orbit brings it sufficiently close. The only information is the instantaneous projected distance of about $1600$ AU which constrains the semi-major axis but provides no information on the eccentricity. At high eccentricity, the Kozai-Lidov time is shortened by a factor of order $(1-e^2)^{3/2}$. For a semi-major axis of 1600 AU and a high eccentricity $1-e^2 \lesssim 0.1$ the Kozai-Lidov time scale for precession and oscilations of the AB orbit is reduced to values comparable to that of the A1-A2 orbit enabling changes in the mutual inclination of the A1-A2+B triple system as well as possible changes in the eccentricity of the AB orbit \citep[][]{Hamers15} which could then change the conclusion about the possible eccentricity variations of the A1-A2 system. The chance for such a high eccentricity in the AB-D orbit could be substantial. For example, if randomly chosen from a random thermal distribution $dn/de\propto e$, which is likely characteristic of binaries at $\sim 1000$AU separations \citep[e.g.][]{DuquennoyMayor91,Hwang22,Tokovinin20}, the chance for $1-e^2<0.1$ is about 10$\%$. A higher eccentricity $1-e^2<0.05$, (a $5\%$ chance for a thermal distribution), would bring the D system to within 40 AU of the triple which is sufficiently close to allow energy exchange and stochastic semi-major axis evolution of the AB-D orbit over long times \citep[e.g.][]{Heggie75,RoyHaddow03,Mushkin20}. Without additional information it is thus not possible to determine the dynamical evolution of the system. An accurate determination of the relative proper motion of the AB-D system, which will hopefully be achieved by future observations (in particular future Gaia releases) may allow useful constraints.

% There is negligible tidal dissipation

%The situation may change dramatically if the distant orbi

%Excitation of eccentricity result in orbital shrinkage due to tidal dissipation \citep[e.g.][]{Kiseleva98,Fabrycky07,Naoz14} or in extreme cases in the merger or collision of the components \citep{Thompson11,Katz12}.
%For a triple system of comparable masses with an inner binary in orbit with a distant mass the Kozai-Lidov time scale for angular momentum exchange is roughly given by
%\begin{equation}
%\tau_{KL}\sim (1-e_{\mathrm{out})^{-3/2}\frac{P_{\mathrm{out}}^2}{P_{\mathrm{in}}}
%\end{equation}
%\noindent where $P_{\mathrm{in}}$ and  $P_{\mathrm{out}}$ are the periods of the inner and outer orbits \citep[e.g.][]{Soderhjelm82}. The Kozai-Lidov time fot the A1-A2 binary is about a hundred thousand years. 

%The time scale for the 

%In the case of HIP 87813, $\tau_{KL} \sim 0.4 \text{ Myrs}$ is much shorter than the age of the system and therefore quite viable to have operated long enough to cause the migration. 
%If the eccentricity is sufficiently high ($e\gtrsim 0.9)$ which we will henceforth assume is not very high ($e\lesssim 0.9$). We note that the possibil

 %- Kozai of the inner triple: time scale: o.k. , GR +  inclination => no Kozai
 %- Tides are likely not enough to cause significant migration. 
 %- Effect of the outer binary significant kick possible each orbital time. Limitations beyond the scope of this paper
 
\subsection{X-ray emission}

HIP 87813 was detected as an X-ray source by ROSAT in the catalogue of pointed observations with the High Resolution Imager \citep[HRI;][]{ROSAT00} in an observation between 1994-09-16 and 26. The count rate of $0.0178 \pm 0.0026 \text{ photons} \text{ s}^{-1}$ within $0.1-2.4$ keV was converted to an X-ray luminosity $L_X \sim 0.95 \times 10^{30} \text{ erg} \text{ s}^{-1}$ in the catalogue of X-ray emitting A stars of \cite{Schroder07}. By comparing the reported count rate and X-ray luminosity to other stars at various distances, we noticed that the reported X-ray luminosity for HIP 87813 is exactly a factor of 10 higher than what it should be (most likely due to a typo), so that its actual X-ray luminosity is $10^{29} \text{ erg} \text{ s}^{-1}$. 

Because of the positional error of 1.9" on the X-ray source, we cannot absolutely exclude that X-ray emission is due to the faint background source of unknown nature. However, assuming that the X-ray emission is from HIP 87813, it is mostly probably associated with one or both the late type stars because A stars have neither a convective envelope nor a strong wind (and are therefore the most X-ray inactive class of MS stars). An X-ray luminosity of $10^{29} \text{ erg} \text{ s}^{-1}$ is fully compatible with that expected from early-mid K stars with an age of about 200 Myrs \citep{Garces11}, and a rotation period on the order of 10 days \citep{Pizzolato03}.

\subsection{How did HJ2814 form and how will it evolve?}
A study of the formation and future evolution of this systems is beyond the scope of this paper. Below we briefly outline several interesting possibilities which we cannot rule out at this point.

The triple A1-A2+B and the binary D1-D2, which are separated by around 1600 AU, may have formed at that separation \citep[e.g.][]{Pineda15,Tokovinin17}. Alternatively, the entire system formed at a scale of $\sim 100$ AU, and the D1-D2 binary dynamically migrated outwards due to the multi body gravitational interactions \citep[e.g.][]{ReipurthMikkola12,Mushkin20}. The A-B system is separated by 24 AU, and therefore could have formed by disk fragmentation. The small separation of the inner A1-A2 binary (0.16 AU) is well below the 10 AU limit above which binaries are believed to form \textit{in situ} \citep[e.g.][]{Bate09} and suggests that it formed at a larger separation. Inward migration may have occurred due to interactions with the disk \citep[e.g.][]{Artymowicz91} or through tidal dissipation enhanced by Kozai-Lidov oscilations \citep[e.g.][]{Kiseleva98,Fabrycky07,Naoz14}. As discussed above, tidal dissipation is unlikely to play a role unless the the AB-D
orbit has a high eccentricity with a sufficiently small pericenter, allowing the D1-D2 binary to perturb the A1-A2+B triple. A high eccentricity is expected if the D1-D2 migrated to its large separation due to energy exchanges with the triple. Hence the questions of origin of the large AB-D separation and of the tight A1-A2 are related.

All 5 stars are massive enough to evolve off the main sequence within about a Hubble time. Given the uncertainty in the dynamical effects of the other stars in the system, various outcomes may result for the close binary - a single WD (involving a merger before, during or after the evolution of the A-star), a close MS-WD binary, a close double WD (involving mass transfer that will shorten the MS lifetime of A$_2$), or even a thermonuclear explosion due to the merger or collision of the possible two WDs. Non trivial evolution is all but guaranteed given that even if nothing else is changed, the A star will expand and fill its Roche lobe reaching a radius of about $15 R_{\odot}$ leading to mass transfer. 

An important question is to what extent the system will remain bound following the stellar evolution of the A1-A2 system given the expected mass loss when forming the WD/s. There is evidence that WDs obtain natal kicks when forming which are not larger than 1 km/s \citep{El-Badry18} which may unbind the HJ2814A-HJ2814B system but not the HJ2814A system itself. However, given the likely asymmetric mass-loss associated with the A1-A2 binary evolution, larger kicks cannot be ruled out. It is therefore possible that the resulting system will remain in a binary or quintuple system. Given that the stars of HJ2814B can evolve off the main sequence within a Hubble time, an end state with four or five WDs in the same system is possible. Whatever the consequence of such evolution is, similar evolved systems likely exist within the vicinity of the sun and motivates the challenging study of multiplicity of nearby WDs. The complex case of HJ2814 helps to demonstrate the remaining observational challenge as well as the rapidly growing set of observational and theoretical tools in the ongoing study of multiplicity of stars. 

\begin{acknowledgements}
This work has made use of data from the European Space Agency (ESA) mission Gaia (https://www.cosmos.esa.int/gaia), processed by the Gaia Data Processing and Analysis Consortium (DPAC, https://www.cosmos.esa.int/web/gaia/dpac/consortium). Funding for the DPAC has been provided by national institutions, in particular the institutions participating in the Gaia Multilateral Agreement. This publication makes use of data products from the Two Micron All Sky Survey, which is a joint project of the University of Massachusetts and the Infrared Processing and Analysis Center/California Institute of Technology, funded by the National Aeronautics and Space Administration and the National Science Foundation. This research has made use of the Jean-Marie Mariotti Center \texttt{SearchCal} service \footnote{Available at http://www.jmmc.fr/searchcal} co-developped by LAGRANGE and IPAG. This reaseach has made use of the CDS Astronomical Databases SIMBAD and VIZIER \footnote{Available at http://cdsweb.u-strasbg.fr/},  NASA's Astrophysics Data System Bibliographic Services, NumPy \citep{van2011numpy} and matplotlib, a Python library for publication quality graphics \citep{Hunter2007}.
\end{acknowledgements}

\bibliographystyle{aa}
\bibliography{main.bib}

\begin{thebibliography}{89}
\expandafter\ifx\csname natexlab\endcsname\relax\def\natexlab#1{#1}\fi

\bibitem[{{Absil} {et~al.}(2011){Absil}, {Le Bouquin}, {Berger}, {Lagrange},
  {Chauvin}, {Lazareff}, {Zins}, {Haguenauer}, {Jocou}, {Kern}, {Millan-Gabet},
  {Rochat}, \& {Traub}}]{Absil11}
{Absil}, O., {Le Bouquin}, J.~B., {Berger}, J.~P., {et~al.} 2011, \aap, 535,
  A68

\bibitem[{{Artymowicz} {et~al.}(1991){Artymowicz}, {Clarke}, {Lubow}, \&
  {Pringle}}]{Artymowicz91}
{Artymowicz}, P., {Clarke}, C.~J., {Lubow}, S.~H., \& {Pringle}, J.~E. 1991,
  \apjl, 370, L35

\bibitem[{{Bailer-Jones} {et~al.}(2021){Bailer-Jones}, {Rybizki}, {Fouesneau},
  {Demleitner}, \& {Andrae}}]{Bailer-Jones21}
{Bailer-Jones}, C.~A.~L., {Rybizki}, J., {Fouesneau}, M., {Demleitner}, M., \&
  {Andrae}, R. 2021, \aj, 161, 147

\bibitem[{{Baraffe} {et~al.}(1998){Baraffe}, {Chabrier}, {Allard}, \&
  {Hauschildt}}]{Baraffe98}
{Baraffe}, I., {Chabrier}, G., {Allard}, F., \& {Hauschildt}, P.~H. 1998, \aap,
  337, 403

\bibitem[{{Bate}(2009)}]{Bate09}
{Bate}, M.~R. 2009, \mnras, 392, 1363

\bibitem[{{Belokurov} {et~al.}(2020){Belokurov}, {Penoyre}, {Oh}, {Iorio},
  {Hodgkin}, {Evans}, {Everall}, {Koposov}, {Tout}, {Izzard}, {Clarke}, \&
  {Brown}}]{Belokurov20}
{Belokurov}, V., {Penoyre}, Z., {Oh}, S., {et~al.} 2020, \mnras, 496, 1922

\bibitem[{{Bourges} {et~al.}(2017){Bourges}, {Mella}, {Lafrasse}, {Duvert},
  {Chelli}, {Le Bouquin}, {Delfosse}, \& {Chesneau}}]{Bourges17}
{Bourges}, L., {Mella}, G., {Lafrasse}, S., {et~al.} 2017, VizieR Online Data
  Catalog, II/346

\bibitem[{{Brandt}(2018)}]{Brandt18}
{Brandt}, T.~D. 2018, \apjs, 239, 31

\bibitem[{{Brandt}(2021)}]{Brandt21}
{Brandt}, T.~D. 2021, \apjs, 254, 42

\bibitem[{{Bystrov} {et~al.}(1994){Bystrov}, {Polojentsev}, {Potter},
  {Yagudin}, {Zallez}, \& {Zelaya}}]{Bystrov94}
{Bystrov}, N.~F., {Polojentsev}, D.~D., {Potter}, H.~I., {et~al.} 1994,
  Bulletin d'Information du Centre de Donnees Stellaires, 44, 3

\bibitem[{{Cushing} {et~al.}(2004){Cushing}, {Vacca}, \& {Rayner}}]{Cushing04}
{Cushing}, M.~C., {Vacca}, W.~D., \& {Rayner}, J.~T. 2004, \pasp, 116, 362

\bibitem[{{De Rosa} {et~al.}(2014){De Rosa}, {Patience}, {Wilson}, {Schneider},
  {Wiktorowicz}, {Vigan}, {Marois}, {Song}, {Macintosh}, {Graham}, {Doyon},
  {Bessell}, {Thomas}, \& {Lai}}]{DeRosa14}
{De Rosa}, R.~J., {Patience}, J., {Wilson}, P.~A., {et~al.} 2014, \mnras, 437,
  1216

\bibitem[{{Dong} {et~al.}(2014){Dong}, {Katz}, \& {Socrates}}]{Dong14}
{Dong}, S., {Katz}, B., \& {Socrates}, A. 2014, \apjl, 781, L5

\bibitem[{{Duch{\^e}ne} \& {Kraus}(2013)}]{Duchene13}
{Duch{\^e}ne}, G. \& {Kraus}, A. 2013, \araa, 51, 269

\bibitem[{{Duquennoy} \& {Mayor}(1991)}]{DuquennoyMayor91}
{Duquennoy}, A. \& {Mayor}, M. 1991, \aap, 248, 485

\bibitem[{{Eggleton} \& {Tokovinin}(2008)}]{Eggleton08}
{Eggleton}, P.~P. \& {Tokovinin}, A.~A. 2008, \mnras, 389, 869

\bibitem[{{El-Badry} \& {Rix}(2018)}]{El-Badry18}
{El-Badry}, K. \& {Rix}, H.-W. 2018, \mnras, 480, 4884

\bibitem[{{Fabrycky} \& {Tremaine}(2007)}]{Fabrycky07}
{Fabrycky}, D. \& {Tremaine}, S. 2007, \apj, 669, 1298

\bibitem[{{Gaia Collaboration} {et~al.}(2018){Gaia Collaboration}, {Brown},
  {Vallenari}, {Prusti}, {de Bruijne}, {Babusiaux}, {Bailer-Jones}, {Biermann},
  {Evans}, {Eyer}, {Jansen}, {Jordi}, {Klioner}, {Lammers}, {Lindegren},
  {Luri}, {Mignard}, {Panem}, {Pourbaix}, {Randich}, {Sartoretti}, {Siddiqui},
  {Soubiran}, {van Leeuwen}, {Walton}, {Arenou}, {Bastian}, {Cropper},
  {Drimmel}, {Katz}, {Lattanzi}, {Bakker}, {Cacciari}, {Casta{\~n}eda},
  {Chaoul}, {Cheek}, {De Angeli}, {Fabricius}, {Guerra}, {Holl}, {Masana},
  {Messineo}, {Mowlavi}, {Nienartowicz}, {Panuzzo}, {Portell}, {Riello},
  {Seabroke}, {Tanga}, {Th{\'e}venin}, {Gracia-Abril}, {Comoretto},
  {Garcia-Reinaldos}, {Teyssier}, {Altmann}, {Andrae}, {Audard},
  {Bellas-Velidis}, {Benson}, {Berthier}, {Blomme}, {Burgess}, {Busso},
  {Carry}, {Cellino}, {Clementini}, {Clotet}, {Creevey}, {Davidson}, {De
  Ridder}, {Delchambre}, {Dell'Oro}, {Ducourant},
  {Fern{\'a}ndez-Hern{\'a}ndez}, {Fouesneau}, {Fr{\'e}mat}, {Galluccio},
  {Garc{\'\i}a-Torres}, {Gonz{\'a}lez-N{\'u}{\~n}ez}, {Gonz{\'a}lez-Vidal},
  {Gosset}, {Guy}, {Halbwachs}, {Hambly}, {Harrison}, {Hern{\'a}ndez},
  {Hestroffer}, {Hodgkin}, {Hutton}, {Jasniewicz}, {Jean-Antoine-Piccolo},
  {Jordan}, {Korn}, {Krone-Martins}, {Lanzafame}, {Lebzelter}, {L{\"o}ffler},
  {Manteiga}, {Marrese}, {Mart{\'\i}n-Fleitas}, {Moitinho}, {Mora}, {Muinonen},
  {Osinde}, {Pancino}, {Pauwels}, {Petit}, {Recio-Blanco}, {Richards},
  {Rimoldini}, {Robin}, {Sarro}, {Siopis}, {Smith}, {Sozzetti}, {S{\"u}veges},
  {Torra}, {van Reeven}, {Abbas}, {Abreu Aramburu}, {Accart}, {Aerts},
  {Altavilla}, {{\'A}lvarez}, {Alvarez}, {Alves}, {Anderson}, {Andrei},
  {Anglada Varela}, {Antiche}, {Antoja}, {Arcay}, {Astraatmadja}, {Bach},
  {Baker}, {Balaguer-N{\'u}{\~n}ez}, {Balm}, {Barache}, {Barata}, {Barbato},
  {Barblan}, {Barklem}, {Barrado}, {Barros}, {Barstow}, {Bartholom{\'e}
  Mu{\~n}oz}, {Bassilana}, {Becciani}, {Bellazzini}, {Berihuete}, {Bertone},
  {Bianchi}, {Bienaym{\'e}}, {Blanco-Cuaresma}, {Boch}, {Boeche}, {Bombrun},
  {Borrachero}, {Bossini}, {Bouquillon}, {Bourda}, {Bragaglia}, {Bramante},
  {Breddels}, {Bressan}, {Brouillet}, {Br{\"u}semeister}, {Brugaletta},
  {Bucciarelli}, {Burlacu}, {Busonero}, {Butkevich}, {Buzzi}, {Caffau},
  {Cancelliere}, {Cannizzaro}, {Cantat-Gaudin}, {Carballo}, {Carlucci},
  {Carrasco}, {Casamiquela}, {Castellani}, {Castro-Ginard}, {Charlot},
  {Chemin}, {Chiavassa}, {Cocozza}, {Costigan}, {Cowell}, {Crifo}, {Crosta},
  {Crowley}, {Cuypers}, {Dafonte}, {Damerdji}, {Dapergolas}, {David}, {David},
  {de Laverny}, {De Luise}, {De March}, {de Martino}, {de Souza}, {de Torres},
  {Debosscher}, {del Pozo}, {Delbo}, {Delgado}, {Delgado}, {Di Matteo},
  {Diakite}, {Diener}, {Distefano}, {Dolding}, {Drazinos}, {Dur{\'a}n},
  {Edvardsson}, {Enke}, {Eriksson}, {Esquej}, {Eynard Bontemps}, {Fabre},
  {Fabrizio}, {Faigler}, {Falc{\~a}o}, {Farr{\`a}s Casas}, {Federici},
  {Fedorets}, {Fernique}, {Figueras}, {Filippi}, {Findeisen}, {Fonti},
  {Fraile}, {Fraser}, {Fr{\'e}zouls}, {Gai}, {Galleti}, {Garabato},
  {Garc{\'\i}a-Sedano}, {Garofalo}, {Garralda}, {Gavel}, {Gavras}, {Gerssen},
  {Geyer}, {Giacobbe}, {Gilmore}, {Girona}, {Giuffrida}, {Glass}, {Gomes},
  {Granvik}, {Gueguen}, {Guerrier}, {Guiraud}, {Guti{\'e}rrez-S{\'a}nchez},
  {Haigron}, {Hatzidimitriou}, {Hauser}, {Haywood}, {Heiter}, {Helmi}, {Heu},
  {Hilger}, {Hobbs}, {Hofmann}, {Holland}, {Huckle}, {Hypki}, {Icardi},
  {Jan{\ss}en}, {Jevardat de Fombelle}, {Jonker}, {Juh{\'a}sz}, {Julbe},
  {Karampelas}, {Kewley}, {Klar}, {Kochoska}, {Kohley}, {Kolenberg},
  {Kontizas}, {Kontizas}, {Koposov}, {Kordopatis}, {Kostrzewa-Rutkowska},
  {Koubsky}, {Lambert}, {Lanza}, {Lasne}, {Lavigne}, {Le Fustec}, {Le
  Poncin-Lafitte}, {Lebreton}, {Leccia}, {Leclerc}, {Lecoeur-Taibi},
  {Lenhardt}, {Leroux}, {Liao}, {Licata}, {Lindstr{\o}m}, {Lister}, {Livanou},
  {Lobel}, {L{\'o}pez}, {Managau}, {Mann}, {Mantelet}, {Marchal}, {Marchant},
  {Marconi}, {Marinoni}, {Marschalk{\'o}}, {Marshall}, {Martino}, {Marton},
  {Mary}, {Massari}, {Matijevi{\v{c}}}, {Mazeh}, {McMillan}, {Messina},
  {Michalik}, {Millar}, {Molina}, {Molinaro}, {Moln{\'a}r}, {Montegriffo},
  {Mor}, {Morbidelli}, {Morel}, {Morris}, {Mulone}, {Muraveva}, {Musella},
  {Nelemans}, {Nicastro}, {Noval}, {O'Mullane}, {Ord{\'e}novic},
  {Ord{\'o}{\~n}ez-Blanco}, {Osborne}, {Pagani}, {Pagano}, {Pailler},
  {Palacin}, {Palaversa}, {Panahi}, {Pawlak}, {Piersimoni}, {Pineau}, {Plachy},
  {Plum}, {Poggio}, {Poujoulet}, {Pr{\v{s}}a}, {Pulone}, {Racero}, {Ragaini},
  {Rambaux}, {Ramos-Lerate}, {Regibo}, {Reyl{\'e}}, {Riclet}, {Ripepi}, {Riva},
  {Rivard}, {Rixon}, {Roegiers}, {Roelens}, {Romero-G{\'o}mez}, {Rowell},
  {Royer}, {Ruiz-Dern}, {Sadowski}, {Sagrist{\`a} Sell{\'e}s}, {Sahlmann},
  {Salgado}, {Salguero}, {Sanna}, {Santana-Ros}, {Sarasso}, {Savietto},
  {Schultheis}, {Sciacca}, {Segol}, {Segovia}, {S{\'e}gransan}, {Shih},
  {Siltala}, {Silva}, {Smart}, {Smith}, {Solano}, {Solitro}, {Sordo}, {Soria
  Nieto}, {Souchay}, {Spagna}, {Spoto}, {Stampa}, {Steele},
  {Steidelm{\"u}ller}, {Stephenson}, {Stoev}, {Suess}, {Surdej}, {Szabados},
  {Szegedi-Elek}, {Tapiador}, {Taris}, {Tauran}, {Taylor}, {Teixeira},
  {Terrett}, {Teyssandier}, {Thuillot}, {Titarenko}, {Torra Clotet}, {Turon},
  {Ulla}, {Utrilla}, {Uzzi}, {Vaillant}, {Valentini}, {Valette}, {van Elteren},
  {Van Hemelryck}, {van Leeuwen}, {Vaschetto}, {Vecchiato}, {Veljanoski},
  {Viala}, {Vicente}, {Vogt}, {von Essen}, {Voss}, {Votruba}, {Voutsinas},
  {Walmsley}, {Weiler}, {Wertz}, {Wevers}, {Wyrzykowski}, {Yoldas},
  {{\v{Z}}erjal}, {Ziaeepour}, {Zorec}, {Zschocke}, {Zucker}, {Zurbach}, \&
  {Zwitter}}]{Gaia18}
{Gaia Collaboration}, {Brown}, A.~G.~A., {Vallenari}, A., {et~al.} 2018, \aap,
  616, A1

\bibitem[{{Gaia Collaboration} {et~al.}(2021){Gaia Collaboration}, {Brown},
  {Vallenari}, {Prusti}, {de Bruijne}, {Babusiaux}, {Biermann}, {Creevey},
  {Evans}, {Eyer}, {Hutton}, {Jansen}, {Jordi}, {Klioner}, {Lammers},
  {Lindegren}, {Luri}, {Mignard}, {Panem}, {Pourbaix}, {Randich}, {Sartoretti},
  {Soubiran}, {Walton}, {Arenou}, {Bailer-Jones}, {Bastian}, {Cropper},
  {Drimmel}, {Katz}, {Lattanzi}, {van Leeuwen}, {Bakker}, {Cacciari},
  {Casta{\~n}eda}, {De Angeli}, {Ducourant}, {Fabricius}, {Fouesneau},
  {Fr{\'e}mat}, {Guerra}, {Guerrier}, {Guiraud}, {Jean-Antoine Piccolo},
  {Masana}, {Messineo}, {Mowlavi}, {Nicolas}, {Nienartowicz}, {Pailler},
  {Panuzzo}, {Riclet}, {Roux}, {Seabroke}, {Sordo}, {Tanga}, {Th{\'e}venin},
  {Gracia-Abril}, {Portell}, {Teyssier}, {Altmann}, {Andrae}, {Bellas-Velidis},
  {Benson}, {Berthier}, {Blomme}, {Brugaletta}, {Burgess}, {Busso}, {Carry},
  {Cellino}, {Cheek}, {Clementini}, {Damerdji}, {Davidson}, {Delchambre},
  {Dell'Oro}, {Fern{\'a}ndez-Hern{\'a}ndez}, {Galluccio}, {Garc{\'\i}a-Lario},
  {Garcia-Reinaldos}, {Gonz{\'a}lez-N{\'u}{\~n}ez}, {Gosset}, {Haigron},
  {Halbwachs}, {Hambly}, {Harrison}, {Hatzidimitriou}, {Heiter},
  {Hern{\'a}ndez}, {Hestroffer}, {Hodgkin}, {Holl}, {Jan{\ss}en}, {Jevardat de
  Fombelle}, {Jordan}, {Krone-Martins}, {Lanzafame}, {L{\"o}ffler}, {Lorca},
  {Manteiga}, {Marchal}, {Marrese}, {Moitinho}, {Mora}, {Muinonen}, {Osborne},
  {Pancino}, {Pauwels}, {Petit}, {Recio-Blanco}, {Richards}, {Riello},
  {Rimoldini}, {Robin}, {Roegiers}, {Rybizki}, {Sarro}, {Siopis}, {Smith},
  {Sozzetti}, {Ulla}, {Utrilla}, {van Leeuwen}, {van Reeven}, {Abbas}, {Abreu
  Aramburu}, {Accart}, {Aerts}, {Aguado}, {Ajaj}, {Altavilla}, {{\'A}lvarez},
  {{\'A}lvarez Cid-Fuentes}, {Alves}, {Anderson}, {Anglada Varela}, {Antoja},
  {Audard}, {Baines}, {Baker}, {Balaguer-N{\'u}{\~n}ez}, {Balbinot}, {Balog},
  {Barache}, {Barbato}, {Barros}, {Barstow}, {Bartolom{\'e}}, {Bassilana},
  {Bauchet}, {Baudesson-Stella}, {Becciani}, {Bellazzini}, {Bernet}, {Bertone},
  {Bianchi}, {Blanco-Cuaresma}, {Boch}, {Bombrun}, {Bossini}, {Bouquillon},
  {Bragaglia}, {Bramante}, {Breedt}, {Bressan}, {Brouillet}, {Bucciarelli},
  {Burlacu}, {Busonero}, {Butkevich}, {Buzzi}, {Caffau}, {Cancelliere},
  {C{\'a}novas}, {Cantat-Gaudin}, {Carballo}, {Carlucci}, {Carnerero},
  {Carrasco}, {Casamiquela}, {Castellani}, {Castro-Ginard}, {Castro Sampol},
  {Chaoul}, {Charlot}, {Chemin}, {Chiavassa}, {Cioni}, {Comoretto}, {Cooper},
  {Cornez}, {Cowell}, {Crifo}, {Crosta}, {Crowley}, {Dafonte}, {Dapergolas},
  {David}, {David}, {de Laverny}, {De Luise}, {De March}, {De Ridder}, {de
  Souza}, {de Teodoro}, {de Torres}, {del Peloso}, {del Pozo}, {Delbo},
  {Delgado}, {Delgado}, {Delisle}, {Di Matteo}, {Diakite}, {Diener},
  {Distefano}, {Dolding}, {Eappachen}, {Edvardsson}, {Enke}, {Esquej}, {Fabre},
  {Fabrizio}, {Faigler}, {Fedorets}, {Fernique}, {Fienga}, {Figueras},
  {Fouron}, {Fragkoudi}, {Fraile}, {Franke}, {Gai}, {Garabato},
  {Garcia-Gutierrez}, {Garc{\'\i}a-Torres}, {Garofalo}, {Gavras}, {Gerlach},
  {Geyer}, {Giacobbe}, {Gilmore}, {Girona}, {Giuffrida}, {Gomel}, {Gomez},
  {Gonzalez-Santamaria}, {Gonz{\'a}lez-Vidal}, {Granvik},
  {Guti{\'e}rrez-S{\'a}nchez}, {Guy}, {Hauser}, {Haywood}, {Helmi}, {Hidalgo},
  {Hilger}, {H{\l}adczuk}, {Hobbs}, {Holland}, {Huckle}, {Jasniewicz},
  {Jonker}, {Juaristi Campillo}, {Julbe}, {Karbevska}, {Kervella}, {Khanna},
  {Kochoska}, {Kontizas}, {Kordopatis}, {Korn}, {Kostrzewa-Rutkowska},
  {Kruszy{\'n}ska}, {Lambert}, {Lanza}, {Lasne}, {Le Campion}, {Le Fustec},
  {Lebreton}, {Lebzelter}, {Leccia}, {Leclerc}, {Lecoeur-Taibi}, {Liao},
  {Licata}, {Lindstr{\o}m}, {Lister}, {Livanou}, {Lobel}, {Madrero Pardo},
  {Managau}, {Mann}, {Marchant}, {Marconi}, {Marcos Santos}, {Marinoni},
  {Marocco}, {Marshall}, {Martin Polo}, {Mart{\'\i}n-Fleitas}, {Masip},
  {Massari}, {Mastrobuono-Battisti}, {Mazeh}, {McMillan}, {Messina},
  {Michalik}, {Millar}, {Mints}, {Molina}, {Molinaro}, {Moln{\'a}r},
  {Montegriffo}, {Mor}, {Morbidelli}, {Morel}, {Morris}, {Mulone}, {Munoz},
  {Muraveva}, {Murphy}, {Musella}, {Noval}, {Ord{\'e}novic}, {Orr{\`u}},
  {Osinde}, {Pagani}, {Pagano}, {Palaversa}, {Palicio}, {Panahi}, {Pawlak},
  {Pe{\~n}alosa Esteller}, {Penttil{\"a}}, {Piersimoni}, {Pineau}, {Plachy},
  {Plum}, {Poggio}, {Poretti}, {Poujoulet}, {Pr{\v{s}}a}, {Pulone}, {Racero},
  {Ragaini}, {Rainer}, {Raiteri}, {Rambaux}, {Ramos}, {Ramos-Lerate}, {Re
  Fiorentin}, {Regibo}, {Reyl{\'e}}, {Ripepi}, {Riva}, {Rixon}, {Robichon},
  {Robin}, {Roelens}, {Rohrbasser}, {Romero-G{\'o}mez}, {Rowell}, {Royer},
  {Rybicki}, {Sadowski}, {Sagrist{\`a} Sell{\'e}s}, {Sahlmann}, {Salgado},
  {Salguero}, {Samaras}, {Sanchez Gimenez}, {Sanna}, {Santove{\~n}a},
  {Sarasso}, {Schultheis}, {Sciacca}, {Segol}, {Segovia}, {S{\'e}gransan},
  {Semeux}, {Shahaf}, {Siddiqui}, {Siebert}, {Siltala}, {Slezak}, {Smart},
  {Solano}, {Solitro}, {Souami}, {Souchay}, {Spagna}, {Spoto}, {Steele},
  {Steidelm{\"u}ller}, {Stephenson}, {S{\"u}veges}, {Szabados}, {Szegedi-Elek},
  {Taris}, {Tauran}, {Taylor}, {Teixeira}, {Thuillot}, {Tonello}, {Torra},
  {Torra}, {Turon}, {Unger}, {Vaillant}, {van Dillen}, {Vanel}, {Vecchiato},
  {Viala}, {Vicente}, {Voutsinas}, {Weiler}, {Wevers}, {Wyrzykowski}, {Yoldas},
  {Yvard}, {Zhao}, {Zorec}, {Zucker}, {Zurbach}, \& {Zwitter}}]{Gaia21}
{Gaia Collaboration}, {Brown}, A.~G.~A., {Vallenari}, A., {et~al.} 2021, \aap,
  649, A1

\bibitem[{{Gallenne} {et~al.}(2015){Gallenne}, {M{\'e}rand}, {Kervella},
  {Monnier}, {Schaefer}, {Baron}, {Breitfelder}, {Le Bouquin}, {Roettenbacher},
  {Gieren}, {Pietrzy{\'n}ski}, {McAlister}, {ten Brummelaar}, {Sturmann},
  {Sturmann}, {Turner}, {Ridgway}, \& {Kraus}}]{Gallene15}
{Gallenne}, A., {M{\'e}rand}, A., {Kervella}, P., {et~al.} 2015, \aap, 579, A68

\bibitem[{{Gao} {et~al.}(2022){Gao}, {Toonen}, \& {Leigh}}]{Gao22}
{Gao}, Y., {Toonen}, S., \& {Leigh}, N. 2022, arXiv e-prints, arXiv:2203.05357

\bibitem[{{Garc{\'e}s} {et~al.}(2011){Garc{\'e}s}, {Catal{\'a}n}, \&
  {Ribas}}]{Garces11}
{Garc{\'e}s}, A., {Catal{\'a}n}, S., \& {Ribas}, I. 2011, \aap, 531, A7

\bibitem[{{Glebocki} \& {Gnacinski}(2005)}]{Glebocki05}
{Glebocki}, R. \& {Gnacinski}, P. 2005, VizieR Online Data Catalog, III/244

\bibitem[{{Gravity Collaboration} {et~al.}(2017){Gravity Collaboration},
  {Abuter}, {Accardo}, {Amorim}, {Anugu}, {{\'A}vila}, {Azouaoui}, {Benisty},
  {Berger}, {Blind}, {Bonnet}, {Bourget}, {Brandner}, {Brast}, {Buron},
  {Burtscher}, {Cassaing}, {Chapron}, {Choquet}, {Cl{\'e}net}, {Collin},
  {Coud{\'e} Du Foresto}, {de Wit}, {de Zeeuw}, {Deen},
  {Delplancke-Str{\"o}bele}, {Dembet}, {Derie}, {Dexter}, {Duvert}, {Ebert},
  {Eckart}, {Eisenhauer}, {Esselborn}, {F{\'e}dou}, {Finger}, {Garcia}, {Garcia
  Dabo}, {Garcia Lopez}, {Gendron}, {Genzel}, {Gillessen}, {Gonte}, {Gordo},
  {Grould}, {Gr{\"o}zinger}, {Guieu}, {Haguenauer}, {Hans}, {Haubois}, {Haug},
  {Haussmann}, {Henning}, {Hippler}, {Horrobin}, {Huber}, {Hubert}, {Hubin},
  {Hummel}, {Jakob}, {Janssen}, {Jochum}, {Jocou}, {Kaufer}, {Kellner},
  {Kendrew}, {Kern}, {Kervella}, {Kiekebusch}, {Klein}, {Kok}, {Kolb}, {Kulas},
  {Lacour}, {Lapeyr{\`e}re}, {Lazareff}, {Le Bouquin}, {L{\`e}na}, {Lenzen},
  {L{\'e}v{\^e}que}, {Lippa}, {Magnard}, {Mehrgan}, {Mellein}, {M{\'e}rand},
  {Moreno-Ventas}, {Moulin}, {M{\"u}ller}, {M{\"u}ller}, {Neumann}, {Oberti},
  {Ott}, {Pallanca}, {Panduro}, {Pasquini}, {Paumard}, {Percheron}, {Perraut},
  {Perrin}, {Pfl{\"u}ger}, {Pfuhl}, {Phan Duc}, {Plewa}, {Popovic}, {Rabien},
  {Ram{\'\i}rez}, {Ramos}, {Rau}, {Riquelme}, {Rohloff}, {Rousset},
  {Sanchez-Bermudez}, {Scheithauer}, {Sch{\"o}ller}, {Schuhler}, {Spyromilio},
  {Straubmeier}, {Sturm}, {Suarez}, {Tristram}, {Ventura}, {Vincent},
  {Waisberg}, {Wank}, {Weber}, {Wieprecht}, {Wiest}, {Wiezorrek}, {Wittkowski},
  {Woillez}, {Wolff}, {Yazici}, {Ziegler}, \& {Zins}}]{GRAVITY17}
{Gravity Collaboration}, {Abuter}, R., {Accardo}, M., {et~al.} 2017, \aap, 602,
  A94

\bibitem[{{Grenier} {et~al.}(1999){Grenier}, {Burnage}, {Faraggiana},
  {Gerbaldi}, {Delmas}, {G{\'o}mez}, {Sabas}, \& {Sharif}}]{Grenier99}
{Grenier}, S., {Burnage}, R., {Faraggiana}, R., {et~al.} 1999, \aaps, 135, 503

\bibitem[{{Hamers} {et~al.}(2015){Hamers}, {Perets}, {Antonini}, \& {Portegies
  Zwart}}]{Hamers15}
{Hamers}, A.~S., {Perets}, H.~B., {Antonini}, F., \& {Portegies Zwart}, S.~F.
  2015, \mnras, 449, 4221

\bibitem[{{Hamers} {et~al.}(2021){Hamers}, {Rantala}, {Neunteufel}, {Preece},
  \& {Vynatheya}}]{Hamers21}
{Hamers}, A.~S., {Rantala}, A., {Neunteufel}, P., {Preece}, H., \& {Vynatheya},
  P. 2021, \mnras, 502, 4479

\bibitem[{{Harrington}(1968)}]{Harrington68}
{Harrington}, R.~S. 1968, \aj, 73, 190

\bibitem[{{Heggie}(1975)}]{Heggie75}
{Heggie}, D.~C. 1975, \mnras, 173, 729

\bibitem[{{Henry} \& {McCarthy}(1993)}]{Henry93}
{Henry}, T.~J. \& {McCarthy}, Donald~W., J. 1993, \aj, 106, 773

\bibitem[{{Hillebrandt} \& {Niemeyer}(2000)}]{Hillebrandt00}
{Hillebrandt}, W. \& {Niemeyer}, J.~C. 2000, \araa, 38, 191

\bibitem[{Hunter(2007)}]{Hunter2007}
Hunter, J.~D. 2007, Computing In Science \& Engineering, 9, 90

\bibitem[{{Hut}(1982)}]{Hut82}
{Hut}, P. 1982, \aap, 110, 37

\bibitem[{{Hwang} {et~al.}(2022){Hwang}, {Ting}, \& {Zakamska}}]{Hwang22}
{Hwang}, H.-C., {Ting}, Y.-S., \& {Zakamska}, N.~L. 2022, \mnras, 512, 3383

\bibitem[{{Ito} \& {Ohtsuka}(2019)}]{Ito19}
{Ito}, T. \& {Ohtsuka}, K. 2019, Monographs on Environment, Earth and Planets,
  7, 1

\bibitem[{{Katz} \& {Dong}(2012)}]{Katz12}
{Katz}, B. \& {Dong}, S. 2012, arXiv e-prints, arXiv:1211.4584

\bibitem[{{Kaufer} {et~al.}(1999){Kaufer}, {Stahl}, {Tubbesing},
  {N{\o}rregaard}, {Avila}, {Francois}, {Pasquini}, \& {Pizzella}}]{Kaufer99}
{Kaufer}, A., {Stahl}, O., {Tubbesing}, S., {et~al.} 1999, The Messenger, 95, 8

\bibitem[{{Kiseleva} {et~al.}(1998){Kiseleva}, {Eggleton}, \&
  {Mikkola}}]{Kiseleva98}
{Kiseleva}, L.~G., {Eggleton}, P.~P., \& {Mikkola}, S. 1998, \mnras, 300, 292

\bibitem[{{Klein} \& {Katz}(2017)}]{Klein17}
{Klein}, Y.~Y. \& {Katz}, B. 2017, \mnras, 465, L44

\bibitem[{{Kozai}(1962)}]{Kozai62}
{Kozai}, Y. 1962, \aj, 67, 591

\bibitem[{{Kushnir} {et~al.}(2013){Kushnir}, {Katz}, {Dong}, {Livne}, \&
  {Fern{\'a}ndez}}]{Kushnir13}
{Kushnir}, D., {Katz}, B., {Dong}, S., {Livne}, E., \& {Fern{\'a}ndez}, R.
  2013, \apjl, 778, L37

\bibitem[{{Lapeyrere} {et~al.}(2014){Lapeyrere}, {Kervella}, {Lacour},
  {Azouaoui}, {Garcia-Dabo}, {Perrin}, {Eisenhauer}, {Perraut}, {Straubmeier},
  {Amorim}, \& {Brandner}}]{Lapeyrere14}
{Lapeyrere}, V., {Kervella}, P., {Lacour}, S., {et~al.} 2014, in Society of
  Photo-Optical Instrumentation Engineers (SPIE) Conference Series, Vol. 9146,
  Optical and Infrared Interferometry IV, ed. J.~K. {Rajagopal}, M.~J.
  {Creech-Eakman}, \& F.~{Malbet}, 91462D

\bibitem[{{Lasker} {et~al.}(2008){Lasker}, {Lattanzi}, {McLean}, {Bucciarelli},
  {Drimmel}, {Garcia}, {Greene}, {Guglielmetti}, {Hanley}, {Hawkins},
  {Laidler}, {Loomis}, {Meakes}, {Mignani}, {Morbidelli}, {Morrison},
  {Pannunzio}, {Rosenberg}, {Sarasso}, {Smart}, {Spagna}, {Sturch},
  {Volpicelli}, {White}, {Wolfe}, \& {Zacchei}}]{Lasker08}
{Lasker}, B.~M., {Lattanzi}, M.~G., {McLean}, B.~J., {et~al.} 2008, \aj, 136,
  735

\bibitem[{{Le Bouquin} {et~al.}(2011){Le Bouquin}, {Berger}, {Lazareff},
  {Zins}, {Haguenauer}, {Jocou}, {Kern}, {Millan-Gabet}, {Traub}, {Absil},
  {Augereau}, {Benisty}, {Blind}, {Bonfils}, {Bourget}, {Delboulbe},
  {Feautrier}, {Germain}, {Gitton}, {Gillier}, {Kiekebusch}, {Kluska},
  {Knudstrup}, {Labeye}, {Lizon}, {Monin}, {Magnard}, {Malbet}, {Maurel},
  {M{\'e}nard}, {Micallef}, {Michaud}, {Montagnier}, {Morel}, {Moulin},
  {Perraut}, {Popovic}, {Rabou}, {Rochat}, {Rojas}, {Roussel}, {Roux},
  {Stadler}, {Stefl}, {Tatulli}, \& {Ventura}}]{LeBouquin11}
{Le Bouquin}, J.~B., {Berger}, J.~P., {Lazareff}, B., {et~al.} 2011, \aap, 535,
  A67

\bibitem[{{Lenzen} {et~al.}(2003){Lenzen}, {Hartung}, {Brandner}, {Finger},
  {Hubin}, {Lacombe}, {Lagrange}, {Lehnert}, {Moorwood}, \&
  {Mouillet}}]{Lenzen03}
{Lenzen}, R., {Hartung}, M., {Brandner}, W., {et~al.} 2003, in Society of
  Photo-Optical Instrumentation Engineers (SPIE) Conference Series, Vol. 4841,
  Instrument Design and Performance for Optical/Infrared Ground-based
  Telescopes, ed. M.~{Iye} \& A.~F.~M. {Moorwood}, 944--952

\bibitem[{{Lidov}(1962)}]{Lidov62}
{Lidov}, M.~L. 1962, \planss, 9, 719

\bibitem[{{Lidov} \& {Ziglin}(1976)}]{LidovZiglin76}
{Lidov}, M.~L. \& {Ziglin}, S.~L. 1976, Celestial Mechanics, 13, 471

\bibitem[{{Lindegren} {et~al.}(2021){Lindegren}, {Klioner}, {Hern{\'a}ndez},
  {Bombrun}, {Ramos-Lerate}, {Steidelm{\"u}ller}, {Bastian}, {Biermann}, {de
  Torres}, {Gerlach}, {Geyer}, {Hilger}, {Hobbs}, {Lammers}, {McMillan},
  {Stephenson}, {Casta{\~n}eda}, {Davidson}, {Fabricius}, {Gracia-Abril},
  {Portell}, {Rowell}, {Teyssier}, {Torra}, {Bartolom{\'e}}, {Clotet},
  {Garralda}, {Gonz{\'a}lez-Vidal}, {Torra}, {Abbas}, {Altmann}, {Anglada
  Varela}, {Balaguer-N{\'u}{\~n}ez}, {Balog}, {Barache}, {Becciani}, {Bernet},
  {Bertone}, {Bianchi}, {Bouquillon}, {Brown}, {Bucciarelli}, {Busonero},
  {Butkevich}, {Buzzi}, {Cancelliere}, {Carlucci}, {Charlot}, {Cioni},
  {Crosta}, {Crowley}, {del Peloso}, {del Pozo}, {Drimmel}, {Esquej}, {Fienga},
  {Fraile}, {Gai}, {Garcia-Reinaldos}, {Guerra}, {Hambly}, {Hauser},
  {Jan{\ss}en}, {Jordan}, {Kostrzewa-Rutkowska}, {Lattanzi}, {Liao}, {Licata},
  {Lister}, {L{\"o}ffler}, {Marchant}, {Masip}, {Mignard}, {Mints}, {Molina},
  {Mora}, {Morbidelli}, {Murphy}, {Pagani}, {Panuzzo}, {Pe{\~n}alosa Esteller},
  {Poggio}, {Re Fiorentin}, {Riva}, {Sagrist{\`a} Sell{\'e}s}, {Sanchez
  Gimenez}, {Sarasso}, {Sciacca}, {Siddiqui}, {Smart}, {Souami}, {Spagna},
  {Steele}, {Taris}, {Utrilla}, {van Reeven}, \& {Vecchiato}}]{Lindegren21}
{Lindegren}, L., {Klioner}, S.~A., {Hern{\'a}ndez}, J., {et~al.} 2021, \aap,
  649, A2

\bibitem[{{Maoz} {et~al.}(2014){Maoz}, {Mannucci}, \& {Nelemans}}]{Maoz14}
{Maoz}, D., {Mannucci}, F., \& {Nelemans}, G. 2014, \araa, 52, 107

\bibitem[{{Mathieu} {et~al.}(2004){Mathieu}, {Meibom}, \& {Dolan}}]{Mathieu04}
{Mathieu}, R.~D., {Meibom}, S., \& {Dolan}, C.~J. 2004, \apjl, 602, L121

\bibitem[{{Milson} {et~al.}(2020){Milson}, {Barton}, \& {Bennett}}]{Milson20}
{Milson}, N., {Barton}, C., \& {Bennett}, P.~D. 2020, arXiv e-prints,
  arXiv:2011.13914

\bibitem[{{Moe} \& {Di Stefano}(2017)}]{Moe17}
{Moe}, M. \& {Di Stefano}, R. 2017, \apjs, 230, 15

\bibitem[{{Moffat}(1969)}]{Moffat69}
{Moffat}, A.~F.~J. 1969, \aap, 3, 455

\bibitem[{{Murphy} {et~al.}(2018){Murphy}, {Moe}, {Kurtz}, {Bedding},
  {Shibahashi}, \& {Boffin}}]{Murphy18}
{Murphy}, S.~J., {Moe}, M., {Kurtz}, D.~W., {et~al.} 2018, \mnras, 474, 4322

\bibitem[{{Mushkin} \& {Katz}(2020)}]{Mushkin20}
{Mushkin}, J. \& {Katz}, B. 2020, \mnras, 498, 665

\bibitem[{{Naoz}(2016)}]{Naoz16}
{Naoz}, S. 2016, \araa, 54, 441

\bibitem[{{Naoz} \& {Fabrycky}(2014)}]{Naoz14}
{Naoz}, S. \& {Fabrycky}, D.~C. 2014, \apj, 793, 137

\bibitem[{{Nordstrom} \& {Andersen}(1985)}]{Nordstrom85}
{Nordstrom}, B. \& {Andersen}, J. 1985, \aaps, 61, 53

\bibitem[{{Perets} \& {Fabrycky}(2009)}]{Perets09}
{Perets}, H.~B. \& {Fabrycky}, D.~C. 2009, \apj, 697, 1048

\bibitem[{{Pineda} {et~al.}(2015){Pineda}, {Offner}, {Parker}, {Arce},
  {Goodman}, {Caselli}, {Fuller}, {Bourke}, \& {Corder}}]{Pineda15}
{Pineda}, J.~E., {Offner}, S. S.~R., {Parker}, R.~J., {et~al.} 2015, \nat, 518,
  213

\bibitem[{{Pizzolato} {et~al.}(2003){Pizzolato}, {Maggio}, {Micela},
  {Sciortino}, \& {Ventura}}]{Pizzolato03}
{Pizzolato}, N., {Maggio}, A., {Micela}, G., {Sciortino}, S., \& {Ventura}, P.
  2003, \aap, 397, 147

\bibitem[{{Raghavan} {et~al.}(2010){Raghavan}, {McAlister}, {Henry}, {Latham},
  {Marcy}, {Mason}, {Gies}, {White}, \& {ten Brummelaar}}]{Raghavan10}
{Raghavan}, D., {McAlister}, H.~A., {Henry}, T.~J., {et~al.} 2010, \apjs, 190,
  1

\bibitem[{{Rayner} {et~al.}(2003){Rayner}, {Toomey}, {Onaka}, {Denault},
  {Stahlberger}, {Vacca}, {Cushing}, \& {Wang}}]{Rayner03}
{Rayner}, J.~T., {Toomey}, D.~W., {Onaka}, P.~M., {et~al.} 2003, \pasp, 115,
  362

\bibitem[{{Reipurth} \& {Mikkola}(2012)}]{ReipurthMikkola12}
{Reipurth}, B. \& {Mikkola}, S. 2012, \nat, 492, 221

\bibitem[{{ROSAT Scientific Team}(2000)}]{ROSAT00}
{ROSAT Scientific Team}. 2000, VizieR Online Data Catalog, IX/28A

\bibitem[{{Rousset} {et~al.}(2003){Rousset}, {Lacombe}, {Puget}, {Hubin},
  {Gendron}, {Fusco}, {Arsenault}, {Charton}, {Feautrier}, {Gigan}, {Kern},
  {Lagrange}, {Madec}, {Mouillet}, {Rabaud}, {Rabou}, {Stadler}, \&
  {Zins}}]{Rousset03}
{Rousset}, G., {Lacombe}, F., {Puget}, P., {et~al.} 2003, in Society of
  Photo-Optical Instrumentation Engineers (SPIE) Conference Series, Vol. 4839,
  Adaptive Optical System Technologies II, ed. P.~L. {Wizinowich} \&
  D.~{Bonaccini}, 140--149

\bibitem[{{Roy} \& {Haddow}(2003)}]{RoyHaddow03}
{Roy}, A. \& {Haddow}, M. 2003, Celestial Mechanics and Dynamical Astronomy,
  87, 411

\bibitem[{{Royer} {et~al.}(2007){Royer}, {Zorec}, \& {G{\'o}mez}}]{Royer07}
{Royer}, F., {Zorec}, J., \& {G{\'o}mez}, A.~E. 2007, \aap, 463, 671

\bibitem[{{Sana} {et~al.}(2014){Sana}, {Le Bouquin}, {Lacour}, {Berger},
  {Duvert}, {Gauchet}, {Norris}, {Olofsson}, {Pickel}, {Zins}, {Absil}, {de
  Koter}, {Kratter}, {Schnurr}, \& {Zinnecker}}]{Sana14}
{Sana}, H., {Le Bouquin}, J.~B., {Lacour}, S., {et~al.} 2014, \apjs, 215, 15

\bibitem[{{Schr{\"o}der} \& {Schmitt}(2007)}]{Schroder07}
{Schr{\"o}der}, C. \& {Schmitt}, J.~H.~M.~M. 2007, \aap, 475, 677

\bibitem[{{Siess} {et~al.}(2000){Siess}, {Dufour}, \& {Forestini}}]{Siess00}
{Siess}, L., {Dufour}, E., \& {Forestini}, M. 2000, \aap, 358, 593

\bibitem[{{Skrutskie} {et~al.}(2006){Skrutskie}, {Cutri}, {Stiening},
  {Weinberg}, {Schneider}, {Carpenter}, {Beichman}, {Capps}, {Chester},
  {Elias}, {Huchra}, {Liebert}, {Lonsdale}, {Monet}, {Price}, {Seitzer},
  {Jarrett}, {Kirkpatrick}, {Gizis}, {Howard}, {Evans}, {Fowler}, {Fullmer},
  {Hurt}, {Light}, {Kopan}, {Marsh}, {McCallon}, {Tam}, {Van Dyk}, \&
  {Wheelock}}]{Skrutskie06}
{Skrutskie}, M.~F., {Cutri}, R.~M., {Stiening}, R., {et~al.} 2006, \aj, 131,
  1163

\bibitem[{{Thompson}(2011)}]{Thompson11}
{Thompson}, T.~A. 2011, \apj, 741, 82

\bibitem[{{Tokovinin}(2014)}]{Tokovinin14}
{Tokovinin}, A. 2014, \aj, 147, 87

\bibitem[{{Tokovinin}(2017)}]{Tokovinin17}
{Tokovinin}, A. 2017, \mnras, 468, 3461

\bibitem[{{Tokovinin}(2018)}]{Tokovinin18}
{Tokovinin}, A. 2018, \apjs, 235, 6

\bibitem[{{Tokovinin}(2019)}]{Tokovinin19}
{Tokovinin}, A. 2019, \aj, 158, 222

\bibitem[{{Tokovinin}(2020)}]{Tokovinin20}
{Tokovinin}, A. 2020, \mnras, 496, 987

\bibitem[{{Tokovinin} {et~al.}(2006){Tokovinin}, {Thomas}, {Sterzik}, \&
  {Udry}}]{Tokovinin06}
{Tokovinin}, A., {Thomas}, S., {Sterzik}, M., \& {Udry}, S. 2006, \aap, 450,
  681

\bibitem[{{Tokovinin}(1997)}]{Tokovinin97}
{Tokovinin}, A.~A. 1997, \aaps, 124, 75

\bibitem[{{Toonen} {et~al.}(2016){Toonen}, {Hamers}, \& {Portegies
  Zwart}}]{Toonen16}
{Toonen}, S., {Hamers}, A., \& {Portegies Zwart}, S. 2016, Computational
  Astrophysics and Cosmology, 3, 6

\bibitem[{Van Der~Walt {et~al.}(2011)Van Der~Walt, Colbert, \&
  Varoquaux}]{van2011numpy}
Van Der~Walt, S., Colbert, S.~C., \& Varoquaux, G. 2011, Computing in Science
  \& Engineering, 13, 22

\bibitem[{{van Leeuwen}(2007)}]{vanLeeuwen07}
{van Leeuwen}, F. 2007, \aap, 474, 653

\bibitem[{{Waisberg}(2019)}]{Waisberg19}
{Waisberg}, I.~R. 2019, PhD thesis, LMU Munich, Germany

\bibitem[{{Woillez} {et~al.}(2019){Woillez}, {Abad}, {Abuter}, {Aller
  Carpentier}, {Alonso}, {Andolfato}, {Barriga}, {Berger}, {Beuzit}, {Bonnet},
  {Bourdarot}, {Bourget}, {Brast}, {Caniguante}, {Cottalorda}, {Darr{\'e}},
  {Delabre}, {Delboulb{\'e}}, {Delplancke-Str{\"o}bele}, {Dembet}, {Donaldson},
  {Dorn}, {Dupeyron}, {Dupuy}, {Egner}, {Eisenhauer}, {Fischer}, {Frank},
  {Fuenteseca}, {Gitton}, {Gont{\'e}}, {Guerlet}, {Guieu}, {Gutierrez},
  {Haguenauer}, {Haimerl}, {Haubois}, {Heritier}, {Huber}, {Hubin}, {Jolley},
  {Jocou}, {Kirchbauer}, {Kolb}, {Kosmalski}, {Krempl}, {Le Bouquin}, {Le
  Louarn}, {Lilley}, {Lopez}, {Magnard}, {Mclay}, {Meilland}, {Meister},
  {Merand}, {Moulin}, {Pasquini}, {Paufique}, {Percheron}, {Pettazzi}, {Pfuhl},
  {Phan}, {Pirani}, {Quentin}, {Rakich}, {Ridings}, {Riedel}, {Reyes},
  {Rochat}, {Santos Tom{\'a}s}, {Schmid}, {Schuhler}, {Shchekaturov}, {Seidel},
  {Soenke}, {Stadler}, {Stephan}, {Su{\'a}rez}, {Todorovic}, {Valdes},
  {Verinaud}, {Zins}, \& {Z{\'u}{\~n}iga-Fern{\'a}ndez}}]{Woillez19}
{Woillez}, J., {Abad}, J.~A., {Abuter}, R., {et~al.} 2019, \aap, 629, A41

\bibitem[{{Zahn}(1977)}]{Zahn77}
{Zahn}, J.~P. 1977, \aap, 500, 121

\bibitem[{{Zanazzi}(2021)}]{Zanazzi21}
{Zanazzi}, J.~J. 2021, arXiv e-prints, arXiv:2112.05868

\bibitem[{{Zorec} \& {Royer}(2012)}]{Zorec12}
{Zorec}, J. \& {Royer}, F. 2012, \aap, 537, A120

\end{thebibliography}

\begin{appendix}

\section{Tables} 

\begin{table}[t]
\centering
\caption{\label{table:RV_hist} Individual radial velocities from literature.}
\begin{tabular}{ccc}
\hline \hline
\shortstack{Date\\(JD-2440000)} & \shortstack{RV\\(km $\text{s}^{-1}$)} & source \\ [0.3cm]
2675.51 & $-21.0 \pm 2.6$ & \cite{Nordstrom85} \\
3262.82 & $-42.1 \pm 1.4$ & \cite{Nordstrom85} \\
3266.87 & $-9.7 \pm 1.3$  & \cite{Nordstrom85} \\
9615.51 & $3.92 \pm 0.55$ & \cite{Grenier99} \\
\hline
\end{tabular}
\end{table}

\begin{table}[t]
\centering
\caption{\label{table:RV_feros} Individual radial velocities from FEROS.}
\begin{tabular}{cc}
\hline \hline
\shortstack{Date\\(JD-2440000)} & \shortstack{RV\\(km $\text{s}^{-1}$)} \\ [0.3cm]
13516.57 & $-28.28 \pm 0.22$ \\ 
13516.61 & $-28.43 \pm 0.07$ \\ 
13516.66 & $-27.97 \pm 0.04$ \\ 
13516.71 & $-27.44 \pm 0.04$ \\ 
13516.76 & $-27.21 \pm 0.02$ \\ 
13516.82 & $-26.83 \pm 0.04$ \\ 
13516.87 & $-26.37 \pm 0.07$ \\ 
13516.91 & $-26.07 \pm 0.22$ \\ 
13517.61 & $-21.26 \pm 0.06$ \\ 
13517.66 & $-20.73 \pm 0.07$ \\ 
13517.70 & $-20.27 \pm 0.04$ \\ 
13517.76 & $-20.04 \pm 0.04$ \\ 
13517.81 & $-19.51 \pm 0.06$ \\ 
13517.86 & $-19.05 \pm 0.06$ \\ 
13517.90 & $-18.37 \pm 0.06$ \\ 
13518.60 & $-13.34 \pm 0.07$ \\ 
13518.68 & $-12.42 \pm 0.26$ \\ 
13915.49 & $-23.25 \pm 0.13$ \\
13915.53 & $-24.24 \pm 0.07$ \\
13915.57 & $-25.23 \pm 0.07$ \\ 
13915.62 & $-26.37 \pm 0.11$ \\ 
\hline
\end{tabular}
\end{table}

\begin{table}[t]
\centering
\caption{\label{table:RV_spex} Individual radial velocities from SpeX.}
\begin{tabular}{cc}
\hline \hline
\shortstack{Date\\(JD-2440000)} & \shortstack{RV\\(km $\text{s}^{-1}$)} \\ [0.3cm]
17607.87 & $-33 \pm 5$ \\ 
17610.82 & $-21 \pm 5$ \\
17638.72 & $-16 \pm 5$ \\ 
17898.98 & $31  \pm 5$ \\ 
17903.10 & $-44 \pm 5$ \\
17909.12 & $20  \pm 5$ \\
17962.76 & $6   \pm 5$ \\
17976.78 & $9   \pm 5$ \\
17980.83 & $-7  \pm 5$ \\
18009.81 & $-47 \pm 5$ \\
18226.16 & $-31 \pm 5$ \\
18257.98 & $3   \pm 5$ \\
18342.72 & $4   \pm 5$ \\
18374.73 & $-35 \pm 5$ \\
18604.10 & $-38 \pm 5$ \\
18615.98 & $-53 \pm 5$ \\
18617.05 & $-18 \pm 5$ \\
18979.97 & $-21 \pm 5$ \\
19029.05 & $-35 \pm 5$ \\
19032.80 & $-25 \pm 5$ \\
19033.75 & $-26 \pm 5$ \\
\hline
\end{tabular}
\end{table}

\begin{table*}[t]
\centering
\caption{\label{table:positions} Positions of HIP 87813 used in the outer orbit fit.}
\begin{tabular}{cccc}
\hline \hline
Epoch & \shortstack{RA2000\\(deg$\pm$mas)} & \shortstack{DEC2000\\(deg$\pm$mas)} & source \\ [0.3cm]
1991.25 & $269.07933841 \pm 0.34$ & $-15.81236061 \pm 0.25$ & $Hipparcos$ \\ [0.3cm]
2000.00 & $269.079315 \pm 10$ & $-15.812520 \pm 10$ & GSC2.3.2 \\ [0.3cm]
2015.50 & $269.0793072927252 \pm 0.10$ & $-15.812841095549638 \pm 0.09$ & $Gaia$ DR2 \\ [0.3cm]
2016.00 & $269.07930698044726 \pm 0.09$ & $-15.812851220033004 \pm 0.07$ & $Gaia$ eDR3 \\ [0.3cm]
\hline
\end{tabular}
\end{table*}

\end{appendix}

\end{document}